\documentclass[seceq,epsf]{ptptex}
\usepackage{graphicx}

\newcommand{\gsim}
{\mathrel{\mbox{\raisebox{-1.0ex}
    {$\stackrel{\displaystyle >}{\displaystyle \sim}$}}}}
\newcommand{\lsim}
{\mathrel{\mbox{\raisebox{-1.0ex}
    {$\stackrel{\displaystyle <}{\displaystyle \sim}$}}}}

\title{
Phase Structure of Hot and/or Dense QCD \\ 
with the Schwinger-Dyson Equation
}

\author{
Satoshi {\sc Takagi}
\footnote{%
E-mail: satoshi@eken.phys.nagoya-u.ac.jp}
}

\inst{
Department of Physics, Nagoya University, Nagoya 464-8602, 
Japan
}

\recdate{October 16, 2002}

\abst{
We investigate the phase structure of hot
and/or dense QCD
using the Schwinger-Dyson equation (SDE) with the improved ladder
approximation in the Landau gauge.
We show that the phase transition
from the two-flavor
color superconducting (2SC) phase to the quark-gluon plasma 
(QGP) phase is of second order, and that
the scaling properties of the Majorana mass gap and the
diquark condensate are consistent with mean field scaling.
We examine the effect of the antiquark contribution 
and find that
setting the antiquark Majorana mass 
equal to the quark one 
is a good approximation in the medium density region.
We also study the
effect of the Debye screening mass of the gluon
and find that ignoring it causes the critical lines
to move to the region of higher temperature and higher chemical
potential.
}

\begin{document}

\maketitle
\begin{picture}(0,0)(-300,-300)
 \put(0,-20){DPNU-02-11}
\end{picture}
\section{Introduction}
\label{Introduction}

The dynamics of quantum chromodynamics (QCD) 
are very rich, and dynamical chiral symmetry breaking 
is one of the most important
features of QCD. 
In hot and/or dense matter, chiral
symmetry is expected to be restored
(see, e.g., Refs.~\citen{restoration} and \citen{Klevan}).
Furthermore, in recent years, 
there has been great interest in the phenomenon called
``color superconductivity'',
which occurs after chiral symmetry restoration at 
non-zero chemical potential in the low temperature region
(for a recent review, see, e.g., Ref.~\citen{RK}).
Therefore, exploration of the QCD phase diagram, 
including the color superconducting phase, 
is an interesting and important subject 
for the purpose of studying 
not only the mechanism of dynamical symmetry breaking
but also its phenomenological applications in cosmology, the
astrophysics of neutron stars and the physics of heavy ion 
collisions.~\cite{RK,restoration}

The phase structure of QCD at non-zero temperature with
zero chemical potential
has been extensively studied with lattice simulations,
but the simulations at finite chemical potential has begun only
recently and these simulations still involve large errors
[see, e.g., Ref.~\citen{Ka} and references cited therein].
Thus, it is important to investigate 
the phase structure of QCD in the finite temperature 
and/or finite chemical potential region by various other approaches. 

In various non-perturbative approaches, the approach based on  
the Schwinger-Dyson Equation (SDE) is one of the most powerful tools
[for a review, see, e.g., Refs.~\citen{MT} and \citen{Ro}]. 
{}From the SDE with a suitable running coupling 
at zero temperature and zero chemical potential, 
the high energy behavior of the mass function 
is shown to be consistent with the result derived from QCD 
with the operator product expanssion and the renormalization group equation. 
When we use the SDE at non-zero chemical potential,
we can distinguish the 
Majorana mass of the quark from 
that of the antiquark by introducing 
the on-shell projectors of the free quark and the free
antiquark, which are useful especially in the high density region, 
where the antiquarks decouple. 
Furthermore, we can  
include the effect of the long range force mediated by
the magnetic mode of the gluon, 
which may have a substantial effect, 
even in the intermediate 
chemical potential region, as in the high density region.~\cite{Son,Hong}
In the SDE analysis,
the phase structure of QCD in the finite temperature and finite chemical
potential region have been investigated, concentrating on 
the chiral symmetry restoration.~\cite{Tani,Blaschke:1997bj,Ha,Bar,Kir1,Kir2}
In a previous work~\cite{Taka},
solving the SDE with the 
full momentum dependence included, we studied the phase 
transition from the hadron phase to the 
two-flavor color superconducting (2SC) phase
in the region of finite quark chemical potential at zero temperature.
It was shown that the phase transition is of first order and 
the existence of the 2SC phase decreases the critical
chemical potential at which the chiral symmetry is restored. 

In this paper, 
we extend the previous work to non-zero temperature
and solve the coupled SDE for the Majorana masses of the quark
($\Delta^-$)
and antiquark ($\Delta^+$) separately from the SDE for the Dirac mass
($B$) 
in the low and intermediate temperature and chemical potential region. 
The true vacuum is determined by comparing the values 
of the effective potential at these solutions.
In several analyses carried out to this time, the SDE 
without the antiquark Majorana mass ($\Delta^+$)
has been used to estimate the order of 
the Majorana mass gap at intermediate
density ($\mu\sim{300~\rm MeV}$).~\cite{Hong,Shus,pis}
Here, we investigate this antiquark effect in the intermediate
density region
by comparing the results obtained from the SDEs in the following three
cases: 
(case-1)
coupled SDEs for the quark and antiquark Majorana masses;
(case-2) 
SDE for the quark Majorana mass with the antiquark Majorana mass set 
to zero, which is valid in the high density region;~\cite{Hong,Shus,pis} 
(case-3) 
SDE for the Majorana mass with the antiquark Majorana mass set equal 
to the quark one, as in several anlayses carried out using models 
with the contact 4-Fermi
interaction [see, e.g., Refs.~\citen{Be,Ra,Sc,Kit,Kit2,Va}]. 
We solve these three types of SDEs 
for the Majorana masses and discuss the importance of the antiquark
contribution in the region of small chemical potential. 
We find that the antiquark mass is of the same order as the quark mass
in the low and medium density region
($\mu \lsim 600~\mbox{MeV}$, where $\mu$ is the quark chemical
potential),
and setting $\Delta^+=\Delta^-$ is actually a good approximation 
for investigating the phase diagram, the quark Majorana mass gap 
and the diquark condensate.  
Furthermore, we investigate the effect of the Debye screening mass
of the gluon by comparing the results from the SDEs in the case of zero
Debye mass with those in the case of non-zero Debye mass.
We find that omission of the Debye mass of the gluon causes the critical
lines in the phase diagram to move to the region of 
the higher temperature and higher chemical potential.

This paper is organized as follows.
In \S\ref{Preliminary}, we summarize the quark propagator, the gluon
propagator and the running coupling that we use in the present analysis.
Several approximations of the quark propagator are made.
We also give formulas for calculating the chiral condensate 
and the diquark condensate. 
In \S\ref{Effective potential and Schwinger-Dyson equation},
 we present the effective potential for the quark propagator and 
then derive the Schwinger-Dyson equation as a stationary condition of 
the effective potential.
Section~\ref{Numerical analysis} is the main part of the paper,
 where we give the results of the numerical analysis of
 the Schwinger-Dyson equation.
Finally, we give a summary and discussion
 in \S\ref{Summary and Discussion}.
In the appendices we summarize several intricate expressions and formulas.

\section{Preliminaries}
\label{Preliminary}

In this section we present the quark propagator, the gluon
propagator and the running coupling that we use in the numerical
analysis.
Using the imaginary time formalism, we extend the
formalism of Ref.~\citen{Taka} to non-zero temperature
[for the real time formalism, see, e.g., Refs.~\citen{NiKo}].
Below, $p^0$ is related to the 
Matsubara frequency at non-zero temperature $T$ as 
$p^0=(2n+1)\pi{i}T$.
In \S\ref{Nambu-Gorkov fields and quark propagator},
 we introduce the eight-component Majorana spinor
(Nambu-Gorkov field) and give the general form of the full quark
propagator as a matrix in the Nambu-Gorkov space.
The gluon propagator with a screening effect is presented in
 \S\ref{Gluon propagator and the running couplinig}.
We also give the explicit form of the running coupling in the analysis.
We give a formula to calculate the quark-antiquark condensate and the
diquark condensate in \S\ref{Condensates}.

\subsection{Nambu-Gorkov field and quark propagator}
\label{Nambu-Gorkov fields and quark propagator}
In the present analysis,
we regard $u$ and $d$ quarks as massless, but
we consider the current mass of the $s$-quark
to be large enough that the strange quark can be ignored in 
the formation of the diquark condensate; 
we assume that the color superconductivity is realized in the 2-flavor
color superconducting (2SC) phase,~\cite{Al} 
where the color symmetry ${SU}(3)_{c}$ is broken down to its subgroup 
${SU}(2)_{c}$.\footnote{
We believe that the color flavor locked (CFL) phase~\cite{Wilczek}
is realized in the high density region, which is beyond the region of our
analysis, and that the 2SC phase is more stable
than the CFL phase near the chiral restoration point.
}

Because we are interested in the phase structure of QCD,
including the color superconducting phase, it is convenient to use 
the eight-component Majorana spinor (Nambu-Gorkov field)
 instead of the four-component Dirac spinor. The Nambu-Gorkov field is 
expressed as
\begin{eqnarray}
 \Psi=\frac{1}{\sqrt{2}}\left(
			 \begin{array}{c}
			  \psi \\
			  \psi^C 
			 \end{array}\right) \ ,
 \quad \psi^C=C\bar\psi^T \ ,
\label{NambuGorkov}
\end{eqnarray}
where $C=i\gamma^2\gamma^0$.
In the Nambu-Gorkov basis, the
inverse full quark propagator is expressed as~\cite{Taka}
\begin{eqnarray}
\label{inprop}
 iS_{F}(p)^{-1}&=&
 \left(
    \begin{array}{cc}
     (p_0+\mu)\gamma^0-\vec{\gamma}\cdot\vec{p}-B(p) & \Delta(p) \\
     \tilde\Delta(p) & (p_0-\mu)\gamma^0-\vec{\gamma}\cdot\vec{p}-B(-p) 
    \end{array}
  \right) \ , \nonumber\\ 
\end{eqnarray}
\begin{eqnarray}
 &&\hspace{1cm}B(p)^{ab}_{ij}=B_1(p)\delta^{ij}
  (\delta^{ab}-\delta^{a3}\delta^{b3})
  +B_3(p)\delta^{ij}\delta^{a3}\delta^{b3} \ , \nonumber\\
 &&\hspace{1cm}\Delta(p)^{ab}_{ij}=\epsilon_{ij}\epsilon^{ab3}\gamma_5
  [\Delta^{+}(p)\Lambda_p^{+}+\Delta^{-}(p)\Lambda_p^{-}] \ ,\nonumber\\
 &&\hspace{1cm}\tilde\Delta(p)^{ab}_{ij}
  =\gamma^0{\Delta(p)^{\dag}}^{ab}_{ij}\gamma^0 \nonumber\\
 &&\hspace{2.1cm}=-\epsilon_{ij}\epsilon^{ab3}\gamma_5
  [\Delta^{+}(p)\Lambda_p^{-}+\Delta^{-}(p)\Lambda_p^{+}] \ ,\label{massf}
\end{eqnarray}
where $i$ and $j$ are the flavor indices and $a$ and $b$ are 
the color indices.
We have chosen the 3-direction as the direction of color symmetry breaking
in such a way that the quarks carrying the first two colors make a pair.
Here $\Lambda_p^{-}$ and $\Lambda_p^{+}$ are the 
on-shell projectors for the quark and antiquark in the massless limit
[for discussion of the massive case, 
see e.g., Ref.~\citen{Fugleberg:2002rk}]:
\begin{eqnarray}
 \Lambda_p^{\mp}&=&
  \frac{1}{2}\biggl(1\mp\frac{\gamma^0\vec\gamma\cdot\vec{p}}
  {\bar{p}}\biggr) \ .
  \label{projection}
\end{eqnarray}
Note that in our analysis, 
the chemical potential $\mu$ is introduced with respect to the quark number.
Now, the full quark propagator includes four scalar functions,
$B_{1}$ and $B_{3}$, corresponding to the Dirac masses 
responsible for the chiral symmetry breaking,
and $\Delta^{+}$ and $\Delta^{-}$, corresponding to the 
Majorana masses responsible for the color symmetry breaking.
As shown in Refs.~\citen{Ha} and \citen{Taka},  
the Dirac masses $B_{1}$ and $B_{3}$ obey the constraint
\begin{eqnarray}
 \label{diraprop}
  \mbox{Im} \left[ B_{i}(p) \right] = - \mbox{Im}
  \left[ B_{i}(-p) \right] \ , \qquad\ ( i = 1,3 ) 
\end{eqnarray}
while, as shown in Ref.~\citen{Taka}, 
the Majorana masses $\Delta^{+}(p)$ and $\Delta^{-}(p)$ are real and
even functions of $p_0$:
\begin{eqnarray}
 \label{majprop}
  \Delta^{\pm}(p) = \Delta^{\pm}(-p) = 
  \left[ \Delta^{\pm}(p^\ast) \right]^{\ast} \ .
\end{eqnarray}

By taking the inverse of the expression in Eq.~(\ref{inprop}),
the full quark propagator is obtained as
\begin{eqnarray}
 \label{quarkp}
 -iS_{F}(p)&=&\left(
  \begin{array}{l}
   \hspace{0.5cm} R^{-1}_+(p) \hspace{1.7cm} 
    -\{(p_0+\mu)\gamma^0-\vec{\gamma}\cdot\vec{p}-B(p)\}^{-1}
    \Delta(p)R^{-1}_-(p) \\
   -\{(p_0-\mu)\gamma^0-\vec{\gamma}\cdot\vec{p}-B(-p)\}^{-1}
  \tilde\Delta(p)R^{-1}_+(p) \hspace{1.6cm} R^{-1}_-(p) 
  \end{array}
  \right) \ , \nonumber\\
\end{eqnarray}
where
\begin{eqnarray}
 R_+(p)&\equiv&\{(p_0+\mu)\gamma^0-\vec{\gamma}\cdot\vec{p}-B(p)\}
  -\Delta(p)\{(p_0-\mu)\gamma^0-\vec{\gamma}\cdot\vec{p}-B(-p)\}^{-1}
  \tilde\Delta(p) \ ,\nonumber\\
 R_-(p)&\equiv&\{(p_0-\mu)\gamma^0-\vec{\gamma}\cdot\vec{p}-B(-p)\}
  -\tilde\Delta(p)\{(p_0+\mu)\gamma^0-\vec{\gamma}\cdot\vec{p}-B(p)\}^{-1}
  \Delta(p) \ .\nonumber\\
\end{eqnarray}

\subsection{Gluon propagator and running coupling}
\label{Gluon propagator and the running couplinig}

In previous analyses of the phase structure carried out
using the SDE with the improved ladder approximation,
e.g., in Refs.~\citen{Tani} and \citen{Ha},
a gluon propagator with the same form as that at $T=\mu=0$
was used.
However, in a hot and/or dense medium, the gluon
is subject to medium effects and thereby becomes effectively massive. 
In this paper, we include the Debye screening mass of the electric
 mode of the gluon propagator through the hard thermal/dense 
loop approximation and ignore the Meissner masses, 
as done in Ref.~\citen{Taka}.
However, the form of the magnetic mode 
used in, e.g., Refs.~\citen{Son}, \citen{Hong} and \citen{Taka} 
cannot be extrapolated to zero chemical potential
 and zero temperature, because the $k_4^2$-term is omitted.
In the present analysis, therefore, we drop
the Landau damping of the magnetic mode 
and include the $k_4^2$-term. 

It should be noted that the wave function renormalization for 
the quark propagator must be kept equal to 1 to ensure consistency with 
the bare quark-gluon vertex adopted 
in the improved ladder approximation~\cite{Kugo:1992pr}.
At zero temperature and zero chemical potential, 
we can show that the SDE with the ladder approximation 
in the Landau gauge does not result in wave function renormalization
(see, e.g., Ref.~\citen{MT}).
In the present analysis,
therefore, we take the Landau gauge for the gluon propagator,
while assuming wave function renormalization for the quark propagator
to be 1, even at non-zero temperature and/or non-zero 
chemical potential, as was done in Ref.~\citen{Taka}.

The explicit form of the gluon propagator that we use in this paper
is given by
\begin{eqnarray}
 D_{\mu\nu}^{AB}(k)&\equiv&{\delta^{AB}D_{\mu\nu}(k)} \nonumber\\
  &=&i\delta^{AB}\frac{1}{(k_4)^2+\vert\vec{k}\vert^2}
  O^{(1)}_{\mu\nu}
  +i\delta^{AB}\frac{1}{(k_4)^2+\vert\vec{k}\vert^2+2M_D^2}
  O^{(2)}_{\mu\nu} \ , \nonumber\\
\label{gluonp}
\end{eqnarray}
where $k_4 = - i k_0 =2 \pi n T$ (with integer $n$) 
and $M_D$ is the Debye mass of the gluon.
In the hard thermal/dense loop approximation, the Debye mass is given 
by~\cite{Bellac} 
\begin{eqnarray}
 M_D^2&=&
\alpha_s
\biggl[\frac{\pi}{3}\{2N_c+N_f\}{T}^2+\frac{N_f}{\pi}\mu^2\biggr]
 \ ,
\label{Debye mass}
\end{eqnarray}
where $N_c=N_f=3$ and the value of $\alpha_s$ 
is set to the value at an infrared scale, $\alpha_s(E_f)$ 
[see Eqs.~(\ref{run})--(\ref{scale})],
because we are interested in the region in which both the 
temperature and chemical potential are lower than the infrared scale, i.e., 
$T, \mu < E_f$. The quantities $O^{(i)}_{\mu\nu}(i=1,2)$ 
are the polarization tensors, defined by
\begin{eqnarray}
 &&O^{(1)}_{\mu\nu}
=P^\bot_{\mu\nu}+\frac{(u\cdot{k})^2}{(u\cdot{k})^2-k^2}
  P^u_{\mu\nu} \ , \quad 
  O^{(2)}_{\mu\nu}=
 -\frac{(u\cdot{k})^2}{(u\cdot{k})^2-k^2}P^u_{\mu\nu} \ ,
\end{eqnarray}
where
\begin{eqnarray}
 &&P^\bot_{\mu\nu}=g_{\mu\nu}-\frac{k_\mu{k}_\nu}{k^2} \ ,
  \quad 
  P^u_{\mu\nu}=\frac{k_\mu{k}_\nu}{k^2}-\frac{k_\mu{u}_\nu+u_\mu{k}_\nu}
  {u\cdot{k}}+\frac{u_\mu{u}_\nu}{(u\cdot{k})^2}k^2 \ .
\end{eqnarray}
The Lorentz four-vector $u^\mu = (1,0,0,0)$ 
in the gluon propagator reflects the explicit breaking of Lorentz
symmetry due to the existence of a temperature and/or chemical potential 
in the rest frame of the medium.

In the improved ladder approximation~\cite{HM},
the high-energy behavior of the running coupling is
determined by the one-loop renormalization group equation derived in
QCD, and the low-energy behavior is suitably regularized.
As a result, at zero temperature and zero chemical potential,
the high-energy behavior of the mass function derived from 
the SDE with the running coupling is consistent with that derived from 
the operator product expansion (OPE)~\cite{MT}.
In the present analysis, we use the following 
two types of running coupling~\cite{HM,Ao,AKM} 
to check the dependence of the results on the methods for the
regularization of the infrared behavior:
\begin{eqnarray}
\label{run}
 ({\rm I})\qquad &&\alpha_s(E)=\frac{g^2(E)}{4\pi}=
  \frac{12\pi}{11N_c-2N_f}\cdot{f}(t,t_f,t_c) \ , \\ \nonumber\\
&&\hspace{0.5cm}f(t,t_f,t_c)=\left\{
 \begin{array}{ll}
  \frac{1}{t} & \quad \mbox{if} \quad t>t_f, \\
  \frac{1}{t_f}+\frac{(t_f-t_c)^2-(t-t_c)^2}{2t_f^2(t_f-t_c)} &  
   \quad \mbox{if} \quad t_f>t>t_c, \\
  \frac{1}{t_f}+\frac{(t_f-t_c)}{2t_f^2} & \quad \mbox{if} \quad  t_c>t, \\
\end{array}
\right.
\nonumber\\
\label{run2}
 ({\rm II})\qquad &&\alpha_s(E)=\frac{g^2(E)}{4\pi}=
  \frac{12\pi}{11N_c-2N_f}\cdot\frac{1}{\max(t,t_f)} \ , 
\end{eqnarray}
where
\begin{eqnarray}
\label{scale}
 && t=\ln\frac{E^2}{\Lambda_{\rm qcd}^2} \ , \qquad 
  t_f=\ln\frac{E_f^2}{\Lambda_{\rm qcd}^2} \ , \qquad 
  t_c=\ln\frac{E_c^2}{\Lambda_{\rm qcd}^2} \ , \qquad 
\end{eqnarray}
with $E$ being the energy scale, $\Lambda_{\rm qcd}$
 the characteristic scale of QCD,
 and $E_f$ and $E_c$ the infrared cutoff scales introduced
 to regularize the infrared singularity.
Note that the value of the running coupling in the infrared region 
needs to be sufficiently large in order to involve 
the dynamical chiral symmetry breaking.~\cite{HM}
Here we use the running coupling (I) for  $t_f=0.5$ and $t_c=-2.0$, 
around which the various physical 
quantities for the case $T=\mu=0$ are very stable 
with respect to changes in $t_f$.~\cite{Ao}
The running coupling (I) was used in Refs.~\citen{Tani} and \citen{Ha},
while the
running coupling (II) was used in our previous work~\cite{Taka}. 
Therefore, 
we investigate the dependence of the result on the 
regularization methods by adopting several values 
of $t_f$ for both (I) and (II). 
As discussed in the previous subsection,
we assume that the mass of the strange quark is
large enough that we can ignore the $s$-quark in the formation of the diquark 
condensate of $u$ and $d$ quarks.
On the other hand, it is natural to assume that the current mass of 
the $s$-quark is smaller than $\Lambda_{\rm qcd}$. In such a case,
the effect caused by the $s$-quark should be included 
in the running coupling.\footnote{We see later that 
$\Lambda_{\rm qcd}{\sim}{600}$~MeV in the present analysis,
 which is apparently larger than the $s$-quark mass.}
Thus we set $N_f=3$ and $N_c=3$ in the running couplings 
(\ref{run}) and (\ref{run2}). 

\subsection{Condensates}
\label{Condensates}

In this subsection we give formulas to calculate
the chiral condensate and the diquark condensate at non-zero
temperature. 
Because we use the imaginary time formalism,
the formulas for the condensates are obtained by 
replacing the integrations over $p_4$ of the expressions 
at $T=0$ in Ref.~\citen{Taka} with the Matsubara sums as 
$\int\frac{dp_4}{2\pi}\rightarrow{T}\sum_{n}$.
In Ref.~\citen{Taka}, the logarithmic divergence of the integral 
is regularized by introducing the ultraviolet (UV) cutoff $\Lambda_4$ 
for the four-dimensional momentum as $-p^2<\Lambda_4^2$.
In the present analysis,
we introduce the UV cutoff $\Lambda$ for the
spatial momentum as $\vert\vec{p}\vert^2<\Lambda^2$.
Then, the general expression of the chiral condensate is given by
\begin{eqnarray}
 &&\langle{\Omega}\vert\bar\psi_{a}^i\psi^{a}_i(0)\vert{\Omega}
  \rangle_{\Lambda}=-T\sum_{n}
  \int^\Lambda\frac{d^3p}{(2\pi)^3}\mbox{tr}[S_{F11}] \ , 
\label{Conbqq}
\end{eqnarray}
where $\vert{\Omega}\rangle$ is the state at non-zero temperature
and non-zero chemical potential, and the trace is taken in the spinor, 
flavor and color spaces. 
Summations over the color index $a$ and the flavor index $i$ are
implicitly taken on the left-hand side of Eq.~(\ref{Conbqq}).
As in the case $T=0$, 
the formula for the chiral condensate of the charge conjugated quarks
is the same as that of the quarks given above:~\cite{Taka}
$
\langle{\Omega}\vert[\bar\psi_C]_{a}^i[\psi_C]_{i}^a(0)
  \vert{\Omega}\rangle_{\Lambda}
=
\langle{\Omega}\vert\bar\psi_{a}^i\psi_{i}^a(0)
\vert{\Omega}\rangle_{\Lambda}
$.

The general form of the diquark condensate is 
\begin{eqnarray}
 &&\langle{\Omega}\vert(\epsilon^{ij}\epsilon_{ab3})
  [\psi^T]_i^aC\gamma_5\psi_j^b(0)\vert{\Omega}\rangle_{\Lambda}
  =-T\sum_{n}\int^\Lambda\frac{d^3p}{(2\pi)^3}
  \mbox{tr}[\epsilon^{(c)}\epsilon^{(f)}S_{F12}\gamma_5] \ , \nonumber\\ 
\end{eqnarray}
where $\epsilon^{(c)}$ and $\epsilon^{(f)}$ are antisymmetric matrices 
in the color and flavor spaces, respectively:
\begin{eqnarray}
 \{\epsilon^{(c)}\}^{ab}=\epsilon^{ab3} \ , \hspace{1.5cm}
  \{\epsilon^{(f)}\}_{ij}=\epsilon_{ij} \ . 
\end{eqnarray} 
The diquark condensate of the charge conjugated quarks
is same as that of the quarks, except for sign:~\cite{Taka}
$
\langle{\Omega}\vert(\epsilon^{ij}\epsilon_{ab3})
[\psi_C^T]_i^aC\gamma_5[\psi_C]_j^b(0)\vert{\Omega}\rangle_{\Lambda}
=-\langle{\Omega}\vert(\epsilon^{ij}\epsilon_{ab3})
  [\psi^T]_i^aC\gamma_5$
$
\psi_j^b(0)\vert{\Omega}\rangle_{\Lambda}
$.
The explicit forms of the above condensates written 
in terms of $B_i$ and $\Delta^{\pm}$ are given in 
Appendix~\ref{Quark Propagator and the Schwinger-Dyson Equation}.

In the improved ladder approximation at zero temperature and 
zero chemical potential,
the high-energy behavior of the mass function is consistent with that
derived from the OPE.
The chiral condensate calculated from the mass function 
was shown to obey the renormalization group evolution derived
from the OPE (see, e.g., Refs.~\citen{MT}).
Then, we identify the condensates, which are calculated with 
UV cutoff $\Lambda$, with those renormalized at the scale 
$\Lambda$ in QCD.
Therefore, we scale them to the condensates at 1 GeV, 
using the leading renormalization group formulas.
Note that at non-zero temperature, the integral over $p^0$ is converted 
into a Matsubara sum, and 
the cutoff $\Lambda$ is introduced for 
the spatial momentum, not for the four-dimensional 
momentum, as in Ref.~\citen{Taka}. 
Nevertheless, the relations between the condensates 
at the scale $\Lambda$ and those at the scale $E$ have
the same form:
\begin{eqnarray}
 \label{bqqrenom}
 &&\langle{\Omega}\vert\bar\psi_{a}^i\psi^{a}_i(0)\vert{\Omega}
  \rangle_E
  =\biggl[\frac{\alpha_s(\Lambda)}
  {\alpha_s(E)}\biggr]^\kappa
  \langle{\Omega}\vert\bar\psi_{a}^i\psi^{a}_i(0)\vert{\Omega}\rangle_{\Lambda}
  \ , 
\\
 &&\langle{\Omega}\vert(\epsilon^{ij}\epsilon_{ab3}) 
  [\psi^T]_i^aC\psi_j^b(0)\vert{\Omega}\rangle_E
  =\biggl[\frac{\alpha_s(\Lambda)}
  {\alpha_s(E)}\biggr]^{\kappa/2}
  \langle{\Omega}\vert(\epsilon^{ij}\epsilon_{ab3}) 
  [\psi^T]_i^aC\psi_j^b(0)\vert{\Omega}\rangle_\Lambda
  \ ,
\nonumber\\
 \label{qqrenom}
\end{eqnarray}
where
\begin{eqnarray}
 &&\kappa=\frac{9C_2(F)}{11N_c-2N_f}=
  \frac{9}{11N_c-2N_f}\frac{N_c^2-1}{2N_c} \ . 
  \label{AD}
\end{eqnarray}

\section{Effective potential and the Schwinger-Dyson equation}
\label{Effective potential and Schwinger-Dyson equation}

In this section
we present the effective potential for the quark propagator and then
derive the Schwinger-Dyson equation (SDE) as a stationary condition of
the effective potential.

In the present analysis,
we are interested in the differences among the values of the effective 
potential corresponding to the 2SC vacuum, the chiral symmetry
breaking 
($\chi$SB) vacuum and the trivial vacuum.
Then, as explained in \S\ref{Nambu-Gorkov fields and quark propagator},
we assume that the current mass of the strange quark is large enough 
that it can be ignored in the valence quark sector, and 
we include only $u$ and $d$ quarks in the effective action.
In other words, we include the strange quark only as
a sea quark in the present analysis.
Then, 
the effective action for the full quark propagator $S_F$ is given by 
\cite{Corn}
\begin{eqnarray}
\label{Efa}
 \Gamma[S_F]&=&\frac{1}{2}\biggl(-i\mbox{Tr}\mbox{Ln}(S_F^{-1})
 -i\mbox{Tr}({S^{(0)}_F}^{-1}S_F)
  -i\Gamma_{\rm{2PI}}\it[S_F]\biggr) \ , 
\end{eqnarray}
where $\mbox{Tr}$ and $\mbox{Ln}$ are taken in all the spaces and  
$\Gamma_{\rm{2PI}}[S_F]$ represents the contributions 
from the two-particle irreducible (with respect to the quark line)
diagrams.
Here, $S^{(0)}_F$ is the free quark propagator.
The factor 1/2 appears because we use
the eight-component Nambu-Gorkov spinor. 
In the high temperature and high density region, 
the one-gluon exchange approximation is valid, because the coupling is weak.
In the present analysis, we extrapolate this approximation to  
intermediate temperature and intermediate chemical potential, 
and include only the contribution from 
the one-gluon exchange diagram in $\Gamma_{\rm 2PI}[S_F]$:
\begin{eqnarray}
 \Gamma_{\rm 2PI}[S_F]=-\frac{1}{2}
  \mbox{Tr}(S_F\cdot{ig}\Gamma^\mu_A\cdot{S_F}\cdot{ig}\Gamma^\nu_B
  \cdot{D}_{\mu\nu}^{AB}) \ ,
\end{eqnarray} 
where $\Gamma^\mu_A$ is the quark-gluon vertex in the Nambu-Gorkov
basis defined as
\begin{eqnarray}
 \Gamma^\mu_A=
  \left(
   \begin{array}{cc} 
    \gamma^\mu{T}_A & 0 \\
    0 & -\gamma^\mu(T_A)^T
   \end{array}
 \right) \ .
\end{eqnarray}
{}From the effective action (\ref{Efa}), the effective potential in 
the momentum space can be written as
\begin{eqnarray}
 \label{efpm}
 V[S_F]&=&-\Gamma[S_F]/\int{d^4x} \nonumber\\
 &=&\frac{1}{2}T\sum_{n}\int\frac{d^3p}{(2\pi)^3} 
  \biggl(\mbox{ln det}\{S_F(p)\}-\mbox{tr}\{S_F^{(0)-1}(p)S_F(p)\}\biggr)
  \nonumber\\&&
  +\frac{1}{2}T\sum_{n}\int\frac{d^3p}{(2\pi)^3} 
  T\sum_{m}\int\frac{d^3q}{(2\pi)^3}\nonumber\\
 &&\hspace{3cm}
  \frac{1}{2}\mbox{tr}\{S_F(p)\cdot{ig\Gamma_A^\mu}\cdot{S_F(q)}\cdot
  {ig\Gamma_B^\nu}\}\cdot{iD^{AB}_{\mu\nu}(p-q)} \ , \nonumber\\
\end{eqnarray}
where ln, det and tr are taken in the spinor, color and flavor spaces.

The SDE is obtained as 
the stationary condition of the effective potential \\
($\delta{V}[S_F]/\delta{S_F}=0$):
\begin{eqnarray}
 \label{SD}
 S_F^{-1}&=&{S^{(0)}_F}^{-1}-
  (ig\Gamma^\mu_A\cdot{S_F}\cdot{ig}\Gamma^\nu_B)\cdot{D}_{\mu\nu}^{AB} \ .
\end{eqnarray}
In this paper we investigate the effect of the antiquark 
contribution by considering 
the following three
different 
types of the SDEs: 
(case-1) SDE with including the quark and
antiquark Majorana masses properly; 
(case-2) SDE with the 
antiquark Majorana mass omitted, 
as an approximation commonly applied in the high density 
region~\cite{Hong,Shus,pis}; 
(case-3) SDE with the quark Majonara mass 
and the antiquark one set equal, 
as in analyses carried out using models with
the contact 4-Fermi interaction~\cite{Be,Ra,Sc,Kit,Kit2,Va}.   

\vspace{0.2cm}
{\bf \underline{case-1}}
\vspace{0.2cm}

The SDE~(\ref{SD}) leads to the following four coupled equations
for the four scalar functions $B_{1}$, $B_{3}$, $\Delta^{+}$ 
and $\Delta^{-}$:
\begin{eqnarray}
 \label{SD11a}
  B_1(p)&=&T\sum_{n}\int\frac{d^3q}{(2\pi)^3} 
  \frac{1}{2}\pi\alpha_s{D}_{\mu\nu}(q-p) 
  \mbox{tr}[\gamma^{\mu}T^AS_{F11}(q)\gamma^{\nu}T^A\delta^{(c)}_1] \ , 
  \nonumber\\ && \\
 \label{SD11b}
  B_3(p)&=&T\sum_{n}\int\frac{d^3q}{(2\pi)^3} 
  \pi\alpha_s{D}_{\mu\nu}(q-p) 
  \mbox{tr}[\gamma^{\mu}T^AS_{F11}(q)\gamma^{\nu}T^A\delta^{(c)}_3] \ , 
  \nonumber\\
\end{eqnarray}
\begin{eqnarray}
 \label{SD12} 
  \Delta^-(p)&=&T\sum_{n}\int\frac{d^3q}{(2\pi)^3} 
  \frac{1}{2}\pi\alpha_s{D}_{\mu\nu}(q-p) 
  \mbox{tr}[\gamma^{\mu}T^AS_{F12}(q)\gamma^{\nu}(T^A)^T\Lambda_p^-\gamma_5
  \epsilon^{(c)}\epsilon^{(f)}] \ , \nonumber \\ && \\
 \label{SD21} 
  \Delta^+(p)&=&T\sum_{n}\int\frac{d^3q}{(2\pi)^3} 
  \frac{1}{2}\pi\alpha_s{D}_{\mu\nu}(q-p) 
  \mbox{tr}[\gamma^{\mu}T^AS_{F12}(q)\gamma^{\nu}(T^A)^T\Lambda_p^+\gamma_5
  \epsilon^{(c)}\epsilon^{(f)}] \ , \nonumber \\ &&
\label{SD21 b}
\end{eqnarray}
where
\begin{eqnarray}
 \{\delta^{(c)}_1\}^{ab}=\delta^{ab}-\delta^{a3}\delta^{b3} \ , \hspace{1.5cm}
  \{\delta^{(c)}_3\}^{ab}=\delta^{a3}\delta^{b3} \ ,  
\end{eqnarray}
and the traces are taken in the spinor, flavor and color spaces. 
The quantity $\alpha_s=\alpha_s(E)$ on the right-hand sides of
Eqs.~(\ref{SD11a})--(\ref{SD21}) is the running coupling defined in
Eqs.~(\ref{run}) and (\ref{run2}). 
As in Refs.~\citen{Tani}, \citen{Ha} and \citen{Taka}, 
we use the angular averaged form 
$E^2=-p^2-q^2$ as the argument of the running coupling 
for $T>0$ and $\mu>0$ in the present analysis.
The explicit forms of the coupled equations in
Eqs.~(\ref{SD11a})--(\ref{SD21 b}) are given in
Appendix~\ref{Quark Propagator and the Schwinger-Dyson Equation}. 
The SDE with $\Delta^-$ distinguished from $\Delta^+$ at non-zero
$\mu$ is analyzed in Refs.~\citen{Taka} and \citen{Ab}, 
where the analyses are done at zero temperature.

\vspace{0.2cm}
{\bf \underline{case-2}}
\vspace{0.2cm}

The SDE~(\ref{SD}) with 
$\Delta^+=0$ leads to three coupled equations 
for the three scalar functions $B_1$, $B_3$ and
$\Delta^-$.
The equation for $\Delta^-$ is given by 
\begin{eqnarray}
 \label{SD12c2} 
  \Delta^-(p)&=&T\sum_{n}\int\frac{d^3q}{(2\pi)^3} 
  \frac{1}{2}\pi\alpha_s{D}_{\mu\nu}(q-p) 
  \mbox{tr}[\gamma^{\mu}T^AS_{F12}^{(\Delta^+=0)}(q)\gamma^{\nu}(T^A)^T
  \Lambda_p^-\gamma_5
  \epsilon^{(c)}\epsilon^{(f)}] \ . \nonumber \\ &&
\end{eqnarray}
The equations for $B_1$ and $B_3$ 
are given by setting $\Delta^+=0$ in 
Eqs.~(\ref{SD11a}) and (\ref{SD11b}).
The SDE with $\Delta^+=0$ is considered to be valid 
in the extremely high density region 
and is used in many works 
at high density.~\cite{Hong,Shus,pis}

\vspace{0.2cm}
{\bf \underline{case-3}}
\vspace{0.2cm}

The SDE~(\ref{SD}) with $\Delta^-=\Delta^+$ leads to 
three coupled equations for the three scalar functions 
$B_1$, $B_3$ and $\Delta^-(=\Delta^+)$.
The equation for $\Delta^-$ is given by
\begin{eqnarray}
 \label{SD12c3} 
  \Delta^-(p)&=&T\sum_{n}\int\frac{d^3q}{(2\pi)^3} \frac{1}{4}
  \pi\alpha_s{D}_{\mu\nu}(q-p) 
  \mbox{tr}[\gamma^{\mu}T^AS_{F12}^{(\Delta^+=\Delta^-)}(q)
  \gamma^{\nu}(T^A)^T\gamma_5
  \epsilon^{(c)}\epsilon^{(f)}] 
\ .
\nonumber\\
\end{eqnarray}
The equations for $B_1$ and $B_3$ are 
obtained by setting $\Delta^-=\Delta^+$ in
Eqs.~(\ref{SD11a}) and (\ref{SD11b}), respectively.
This type of the Majorana mass gap is used in several 
analyses carried out using models with the 
local 4-Fermi interaction~\cite{Be,Ra,Sc,Kit,Kit2,Va} 
as well as the SDE.~\cite{pSD} 

As we see in Eq.~(\ref{majprop}), the Majonara masses
$\Delta^{+}(p)$ and $\Delta^{-}(p)$ are real, even 
functions of $p_0$,
while in general the Dirac masses $B_{1}(p)$ and $B_{3}(p)$ are 
complex functions.
Equation~(\ref{diraprop}) is the constraint on the imaginary part,
while 
no constraint is obtained on the real part from general
considerations.
However, as shown in Refs.~\citen{Ha} and \citen{Taka} 
for the case $\Delta^{+}(p)=\Delta^{-}(p)=0$, 
the structure of the SDE leads to a natural
constraint on the real parts of the Dirac masses $B_{1}(p)$ and $B_{3}(p)$,
\begin{eqnarray}
\label{diratprop}
B_{1,3}(-p) = B_{1,3}^\ast(p) \ .
\end{eqnarray}
Using the above properties for $B_{1}$ and $B_{3}$ and those 
for $\Delta^{+}$ and $\Delta^{-}$ in Eq.~(\ref{majprop}), 
we can always restrict the summation over the
Matsubara frequencies
to a sum over only the positive frequencies. 

Substituting the solution of Eq.~(\ref{SD}) into Eq.~(\ref{efpm}),
 we obtain the effective potential 
for the vacuum, i.e.~at the stationary point. 
Because the effective potential itself is divergent,
 we subtract the effective potential for the trivial vacuum and define
\begin{eqnarray}
 &&\bar{V}_{\rm sol}[\Delta^+ , \Delta^- , B_1 , B_3]
 \equiv{V}[\Delta^+ , \Delta^- , B_1 , B_3]-V[0 , 0 , 0 , 0] 
\ .
\label{potm}
\end{eqnarray}
The value of the effective potential in Eq.~(\ref{potm}) is understood
as the difference between
the energy density of the vacuum at the stationary point 
and that of the trivial vacuum.
The true vacuum should be determined by evaluating the value
of the effective potential:
The vacuum with the smallest value of 
$\overline{V}_{\rm sol}$ is the most
stable vacuum.
The explicit form of the above expression is given in 
Eq.~(\ref{explicit form of V}) of
Appendix~\ref{Quark Propagator and the Schwinger-Dyson Equation}.

\section{Numerical analysis}
\label{Numerical analysis}

In this section we present the results of the numerical analysis. 
The only parameter necessary to carry out the numerical analysis
is the infrared cutoff parameter $t_f$ in the running coupling.
The unit of the energy scale in the numerical analysis 
is $\Lambda_{\rm qcd}$,  
and it is determined by calculating the 
pion decay constant $f_\pi$ for fixed $t_f$ at zero temperature and 
zero chemical potential through the Pagels-Stokar formula~\cite{Pag},
\begin{eqnarray}
\label{psf}
 f_\pi^2=\frac{N_c}{4\pi^2}\int{p_E^2}{dp_E^2}
  \frac{B(p)\biggl(B(p)-\frac{p_E^2}{2}\frac{dB(p)}
  {dp_E^2}\biggr)}
  {[p_E^2+B^2(p)]^2} \ .
\end{eqnarray}
We use $f_\pi=88$\,MeV, estimated
in the chiral limit~\cite{Gasser:1984yg}, as an input.
At zero temperature and zero chemical potential, 
the dependences of the physical quantities 
on $t_f$ have been shown to be small
around $t_f=0.5$ for the running
coupling (I) with $t_c=-2.0$.~\cite{AKM}
For the running coupling (II), we verified that 
the dependences of the physical quantities 
(mass and condensate) on $t_f$ are small
around $t_f=0.25$.
Therefore, we use
$t_f=0.5$ and $t_c=-2.0$ for the running coupling (I) in
Eq.~(\ref{run})
and $t_f=0.25$ for the running coupling (II) in Eq.~(\ref{run2})
for general $T$ and $\mu$.
{}From these inputs, we obtain $\Lambda_{\rm qcd}=583$ MeV for the
running coupling (I) and $\Lambda_{\rm qcd}=567$ MeV for the running
coupling (II). 
Later in this section we study the dependence of our results on $t_f$.

We introduce the framework of the numerical analysis in
 \S\ref{Framework of the numerical analysis}. 
In \S\ref{Phase Structure} we present the phase diagram obtained from the
 SDE in case-1 and the critical exponents of the mass gaps and condensates
near the phase transition point from the hadronic phase 
to the quark-gluon plasma (QGP) phase 
and that from the 2SC phase to the QGP phase.
We investigate the effect of the Debye mass of the gluon 
in \S\ref{The effect of the Debye mass}.
In \S\ref{antiquark contribution}, we study the effect of the antiquark
contribution by solving the coupled SDEs in the three cases 
introduced in the previous section.

\subsection{Framework of the numerical analysis}
\label{Framework of the numerical analysis}
In this subsection we summarize the framework of the numerical analysis.
First, as discussed below Eq.~(\ref{diratprop}),
 we restrict the Matsubara sum to a sum over the positive frequency
 modes using the properties in 
Eqs.~(\ref{majprop}) and (\ref{diratprop}). 
Second, to solve the SDEs numerically, we transform the variable
 $\bar{p}\equiv\vert\vec{p}\vert$ into the new variable $X$.
For this transformation, we use a $\mu$-independent transformation
 in the region of small chemical potential ($\mu<\mu_0$) and 
a $\mu$-dependent transformation in the region of medium chemical
potential ($\mu\geq\mu_0$).
In the region of small chemical potential,
where the chiral condensate is formed,
the characteristic scale of the system is $\Lambda_{\rm qcd}$.
Here, the dynamical information comes mainly
from the region in which $\bar{p}\lsim\Lambda_{\rm qcd}$.
In the high density region,
 where the diquark condensate is formed, the chemical
 potential $\mu$, in addition to $\Lambda_{\rm qcd}$, gives an important scale.
Here, the dynamically important region is that in which $\bar{p}\sim\mu$. 
Therefore, we make the transformation
from $\bar{p}$ to $X$ as 
\begin{eqnarray}
 \label{des1}
&&
  X = \ln \left[ \bar{p}/\Lambda_{\rm qcd} \right] \ 
 \hspace{1.2cm}\mbox{for} \quad \mu<\mu_0 \ ,
 \qquad \\
 \label{des2}
&&
  X = \ln \left[ \bar{p}/(3\mu) \right] \ 
  \hspace{1.2cm}\mbox{for} \quad \mu\geq\mu_0 \ . 
 \qquad 
\end{eqnarray}
Here, we fix $\mu_0/\Lambda_{\rm qcd}=1/3$, because
the two schemes of the discretization described in
Eqs.~(\ref{des1}) and (\ref{des2}) are connected
continuously to each other at this point.~\cite{Taka} 
We have confirmed that the two schemes
are connected smoothly to each other around $\mu=\mu_0$, 
as in Ref.~\citen{Taka}.

Under the above transformations, 
the integral over $\bar{p}$ on the interval $[0,\infty]$
 is converted into an integral over $X$ on the interval 
$[-\infty,\infty]$. In the numerical integration, 
we introduce ultraviolet (UV) and infrared (IR) cutoffs for $X$,
restricting the region as $X\in[\lambda_{IR}, \lambda_{UV}]$.
We discretize the interval of $X$ by introducing $N_X$ 
evenly spaced points: 
\begin{eqnarray}
 X[J]&=&\lambda_{IR}+\Delta{X}\cdot{J} , \quad J=0,1,\cdots,N_X-1 , 
\end{eqnarray}
where
\begin{eqnarray}
  \Delta{X}=\frac{\lambda_{UV}-\lambda_{IR}}{N_X-1}. \quad 
\end{eqnarray}
The integration over $\bar{p}$ is thus replaced by 
the following summation: 
\begin{eqnarray}
 \int{d\bar{p}} \rightarrow \Delta{X}\sum_{J}e^{X[J]}.
\end{eqnarray}
In the present analysis, for the UV and IR cutoffs, we use 
\begin{eqnarray}
 X: \ [\lambda_{IR},\lambda_{UV}]=[-3.5,2.5] \ .
\label{range}
\end{eqnarray}

For the Matsubara frequency, $p_4=(2n+1)\pi{T}$,
we truncate the infinite sum and perform the 
summation over the $2n_0$ modes labeled by 
integer $n$ in the region $n\in[-n_0, n_0-1]$. 
Throughout this paper, we consider $n_0=40$ and $N_X=40$.
We have checked that 
these values are sufficiently large for the present purpose.
To obtain $\langle\bar\psi\psi\rangle_{\rm 1GeV}$ 
and $\langle\psi\psi\rangle_{\rm 1GeV}$, we use
Eq.~(\ref{bqqrenom}) and Eq.~(\ref{qqrenom}) with 
\begin{eqnarray}
&&\Lambda=\Lambda_{\rm qcd}\exp(\lambda_{UV}) \ 
\hspace{1.3cm}\mbox{for} \quad \mu<\mu_0 \ ,
 \qquad \\
 &&\Lambda=3\mu\exp(\lambda_{UV}) \hspace{0.6cm} 
  \hspace{1.1cm}\mbox{for} \quad \mu\geq\mu_0 \ .
  \qquad 
\end{eqnarray}

We solved the SDEs in case-1 with an iteration method.
Starting from a set of trial functions, we updated the mass functions
with the SDE as
\begin{eqnarray}
 \left\{
  B_{1,3{\rm old}} \,,\, \Delta^{\pm}_{\rm old}
\right\}
\Rightarrow
\fbox{Right-hand sides of SDEs (3.6)--(3.9)}
\Rightarrow
\left\{
  B_{1,3{\rm new}} \,,\, \Delta^{\pm}_{\rm new}
\right\}
\ . \nonumber\\
\end{eqnarray}
Then, we stopped the iteration when the convergence condition
\begin{eqnarray}
 &&\varepsilon\Lambda_{\rm qcd}^6>
  T\sum_{n=-n_0}^{n_0-1}\int^\Lambda\frac{d^3p}{(2\pi)^3}
  \frac{1}{4}{\rm tr}
  \biggl[\biggl(\frac{\delta{V}}{\delta[S_F(p)]}\biggr)^\dag
  \biggl(\frac{\delta{V}}{\delta[S_F(p)]}\biggr)\biggr] \nonumber\\
 &&\hspace{2cm}=T\sum_{n=-n_0}^{n_0-1}\int^\Lambda\frac{d^3p}{(2\pi)^3}
  \biggl\{
  2\vert{B}_{1\rm old}(p)-B_{1\rm new}(p)\vert^2
  +\vert{B}_{3\rm old}(p)-B_{3\rm new}(p)\vert^2
  \nonumber\\
 &&\hspace{3cm}+2\vert\Delta^+_{\rm old}(p)-\Delta^+_{\rm new}(p)\vert^2
  +2\vert\Delta^-_{\rm old}(p)-\Delta^-_{\rm new}(p)\vert^2
  \biggr\}
\end{eqnarray}
was satisfied with suitably small $\varepsilon$.
 In the present analysis, we set $\varepsilon=10^{-10}$. 
In case-2 and case-3, we adopted the same method 
as that used in case-1.

\subsection{Phase structure}
\label{Phase Structure}

\begin{figure}[hb]
 \centerline{\includegraphics[width=12cm]{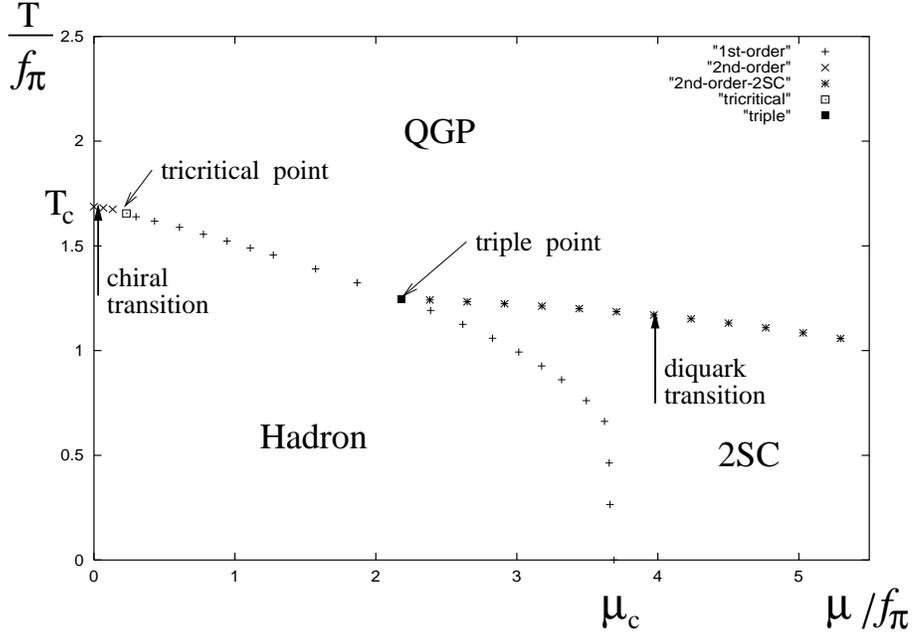}}
 \caption[]{Phase diagram
 for $0\leq{T}/f_\pi\leq{2.5}$ and $0\leq\mu/f_\pi\leq{5.5}$. 
The symbol $\times$ denotes the second order phase transtion
 between the hadron phase and the QGP phase,
$+$ the first order phase transition between the hadron phase
 and the QGP phase or the 2SC phase,
 and $\times\hspace{-0.75em}+$ the second order phase
 transition between the 2SC phase and the QGP phase.
 $T_c$=147 \mbox{MeV}, $\mu_c$=325 \mbox{MeV},
 and ($T$, $\mu$)=(146, 20) \mbox{MeV} at the tricritical point,
 where the second order phase transition changes to 
 first order. At the ``triple point'', ($T$, $\mu$)=(110, 192)
 \mbox{MeV}. Here, the hadron phase 
 ($\langle\bar{\psi}\psi\rangle\neq{0}$), 2SC phase 
 ($\langle\psi\psi\rangle\neq{0}$) and QGP phase 
 ($\langle\bar{\psi}\psi\rangle=\langle\psi\psi\rangle=0$)
 meet. 
 }
 \label{phase-structure}
\end{figure}

\begin{figure}[htb]
 \centerline{
 \includegraphics[width=10cm]{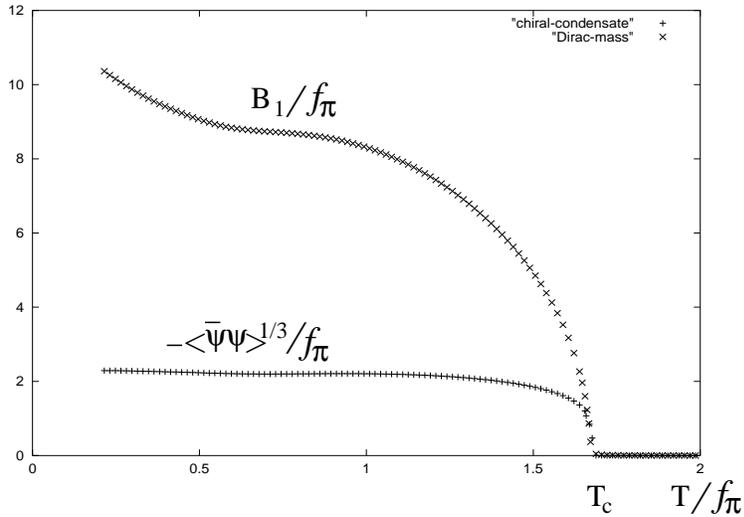}
 }
 \caption[]{
 Temperature dependences 
 of the chiral condensate and the Dirac mass gap at $\mu=0$. 
 Here, the critical temperature of 
 the chiral phase transition from the hadron phase to 
 the QGP phase at $\mu=0$ is $T_c = 147\,\mbox{MeV}$.
}
 \label{chiralcondent}
\end{figure}

Let us first study the phase strucuture by solving the
SDEs~(\ref{SD11a})--(\ref{SD21 b}) in case-1.
The resultant phase diagram 
obtained by using a running coupling of type (I) with $t_f=0.5$ 
and $M_D\neq{0}$ is shown in Fig.~\ref{phase-structure}.
In the present analysis, 
we consider the possibility of three phases:
the hadron ($\chi${SB}) phase, the 2SC phase and the QGP phase.
We determine the true vacuum
by evaluating the value of the effective potential at 
the solution of the SDE. 
The running coupling of type (I) in Eq.~(\ref{run}) used here
becomes very large in the infrared region.
It exceeds the critical value $\pi/2$,\footnote{
  At $T=\mu=0$, the SDE for 
  $\Delta= \Delta^-=\Delta^+$ takes the same
  form as that for the Dirac mass $B$, except that an extra
  factor of $1/2$ appears in front of the integration kernel.
  Because the critical value of the coupling in the SDE for $B$
  is $\pi/4$,~\cite{MT} the extra factor of 1/2
  implies that the critical value of the coupling in the
  SDE for $\Delta$ is $\pi/2$.
}
above which the SDE for the Majonara mass with zero Dirac mass
($B=0$) provides a non-trivial solution even at $T=\mu=0$.
Then, as we have shown in the case $\mu>0$ and $T=0$ 
in Ref.~\citen{Taka},
even for non-zero but small temperature, the 2SC vacuum 
always exits, and in it, a non-trivial Majonara mass 
is dynamically generated,
and the 2SC vacuum is more stable than the trivial vacuum (
the QGP vacuum in the present analysis).
In the low chemical potential region, 
the $\chi$SB vacuum also exists, and in it, the Dirac mass
$B$ is dynamically generated, and it is most stable among
the three vacua.
Hence, the true vacuum in the low temperature and low 
chemical potential region 
is the $\chi$SB vacuum. This is natural, because 
the strength of the attractive force mediated by one-gluon exchange 
between two
quarks in the color anti-triplet channel is 
weaker than that between a quark and
an antiquark in the color singlet channel at $T=\mu=0$. 
When the temperature is increased at $\mu=0$, the value of the
effective potential at the 2SC vacuum goes smoothly to zero 
around $T\sim{110}\,\mbox{MeV}$ ($T/f_\pi\sim{1.3}$), 
while the $\chi$SB vacuum is still the most stable.
Up to this region,
the value of the chiral condensate $\langle\bar{\psi} \psi\rangle$
(per flavor) does not change significantly, as we show 
in Fig.~\ref{chiralcondent}. 
As the temperature is increased further, 
the value of the chiral condensate starts to decrease, and it, 
as well as the mass function in the zero-momentum limit (shown
by $B_1/f_\pi$ in Fig.~\ref{chiralcondent}), finally
goes to zero at $T/f_\pi=1.67$; i.e.,
\begin{equation}
 T_c=147\,\mbox{MeV} \ . \quad (\mu = 0)  
\end{equation}
At the same time, the value of the
effective potential in the $\chi$SB vacuum also 
smoothly goes to zero,
and thus the trivial vacuum appears.
Then, at $\mu=0$ a phase transition occurs from the hadron phase to
the QGP phase (the chiral phase transition), and it is of second
order.
This is consistent with the results of previous analyses performed 
using the SDE (see, e.g., Refs.~\citen{Bar,Tani,Ha,Kir1,Ik}).

We obtain the tricritical point at
\begin{equation}
(T_t, \mu_t)=(146, 20)\,\mbox{MeV} \ ,
\end{equation}
(indicated by {\small $\square$} in Fig.~\ref{phase-structure}): 
For $\mu>\mu_t$ the chiral phase transition is of first order
(indicated by $+$ in Fig.~\ref{phase-structure}), 
while for $\mu<\mu_t$ it is of second order
(indicated by $\times$ in Fig.~\ref{phase-structure}). 
The value of the chemical potential at the tricritical point
obtained in the present analysis is smaller than those in several 
other models, such as the instanton model~\cite{Be},
the NJL model~\cite{Sc,Kit,Kit2} and the random matrix model,~\cite{Va}
as well as in the analysis done using the SDE,~\cite{Bar,Kir1}
but is consistent with that obtained in Ref.~\citen{Ha},
which was also obtained through analysis of the SDE.
We find that in the framework of the SDE,
the large difference in the position of the tricritical
point in the $T$-$\mu$ plane is
caused by the existence of the imaginary part of the Dirac mass. 
We will discuss this point in 
\S\ref{Summary and Discussion}.  

\begin{figure}[htb]
 \centerline{
 \includegraphics[width=10cm]{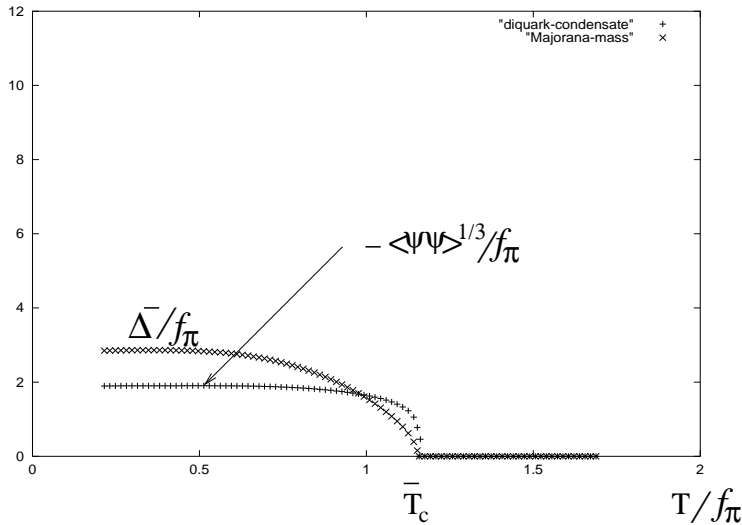}
 }
 \caption[]{
 Temperature dependences 
 of the diquark condensate and the Majorana mass gap at $\mu/f_\pi=4$. 
 Here, the critical temperature of 
 the color phase transition 
 from the 2SC phase to the QGP phase at
 $\mu/f_\pi=4$ is $\bar{T}_c$=102 $\mbox{MeV}$.}
 \label{diquarkcondent}
\end{figure}

When we increase the chemical potential at $T=0$, on the other hand,
the 2SC vacuum becomes more stable than the $\chi$SB vacuum 
at the critical chemical potential $\mu/f_\pi = 3.69$, i.e.,
\begin{equation}
\mu_c=325\,\mbox{MeV} \ , \quad (T=0)
\end{equation}
before the $\chi$SB vacuum becomes less
stable than the trivial vacuum.
As a result, a phase transition occurs from the hadron phase to the
2SC phase, and it is of first order.
This structure is same as that obtained in Ref.~\citen{Taka}, where
the form of the gluon propagator was different from the present one, 
given in Eq.~(\ref{gluonp}), and a running coupling of type (II)
in Eq.~(\ref{run2}) was used.
Let us increase the temperature in the 2SC phase.
The temperature dependences of the diquark condensate
$\langle\psi \psi\rangle$ (per flavor) and the Majonara mass gap of
the quark $\Delta^-$ for $\mu/f_\pi=4$
are shown in Fig.~\ref{diquarkcondent}.
This shows that the diquark condensate and the Majonara mass gap
are stable against change of the
temperature in the low temperature region
($T/f_\pi \lsim 0.6$, i.e., $T\lsim50\,\mbox{MeV}$).
Around $T\sim 50\,\mbox{MeV}$, the condensate and the mass gap 
start to decrease, and they go to zero at
$T=\bar{T}_c = 1.15 f_\pi = 102\,\mbox{MeV}$.
At the same time, the value of the effective potential
in the 2SC vacuum smoothly vanishes.
Then, the phase transition from the 2SC phase to the 
QGP phase occurs, and this color phase transition is of second order. 
This is consistent with the result obtained in 
models with a 4-Fermi interaction~\cite{Be,Kit,Kit2}.
However,
as we can see from Fig.~\ref{phase-structure},
the critical temperature of the color phase transtion 
obtained in the present analsysis 
is around $100\,\mbox{MeV}$
for any $\mu$ satisfying $\mu\lsim500\,\mbox{MeV}$,
and this value is larger than the values obtained in 
models based on the contact 4-Fermi 
interaction~\cite{Be,Sc,Kit,Kit2}.
This increase of the critical temperature may be 
caused by the long range force mediated by the magnetic
mode of the gluon.
We will return to this point in the next subsection.

\begin{figure}[htb]
 \centerline{
 \includegraphics[width=7cm]{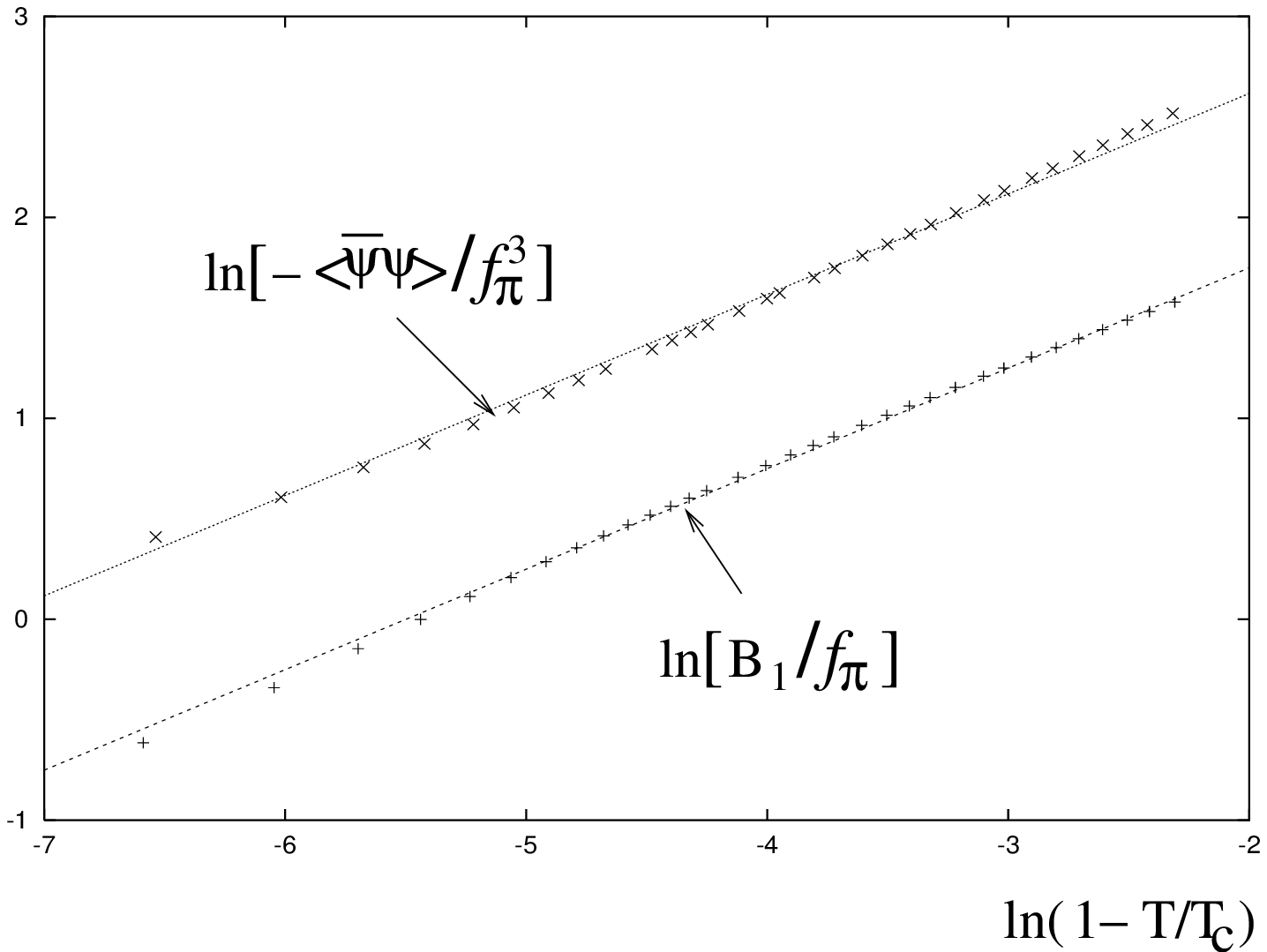}
 \includegraphics[width=7cm]{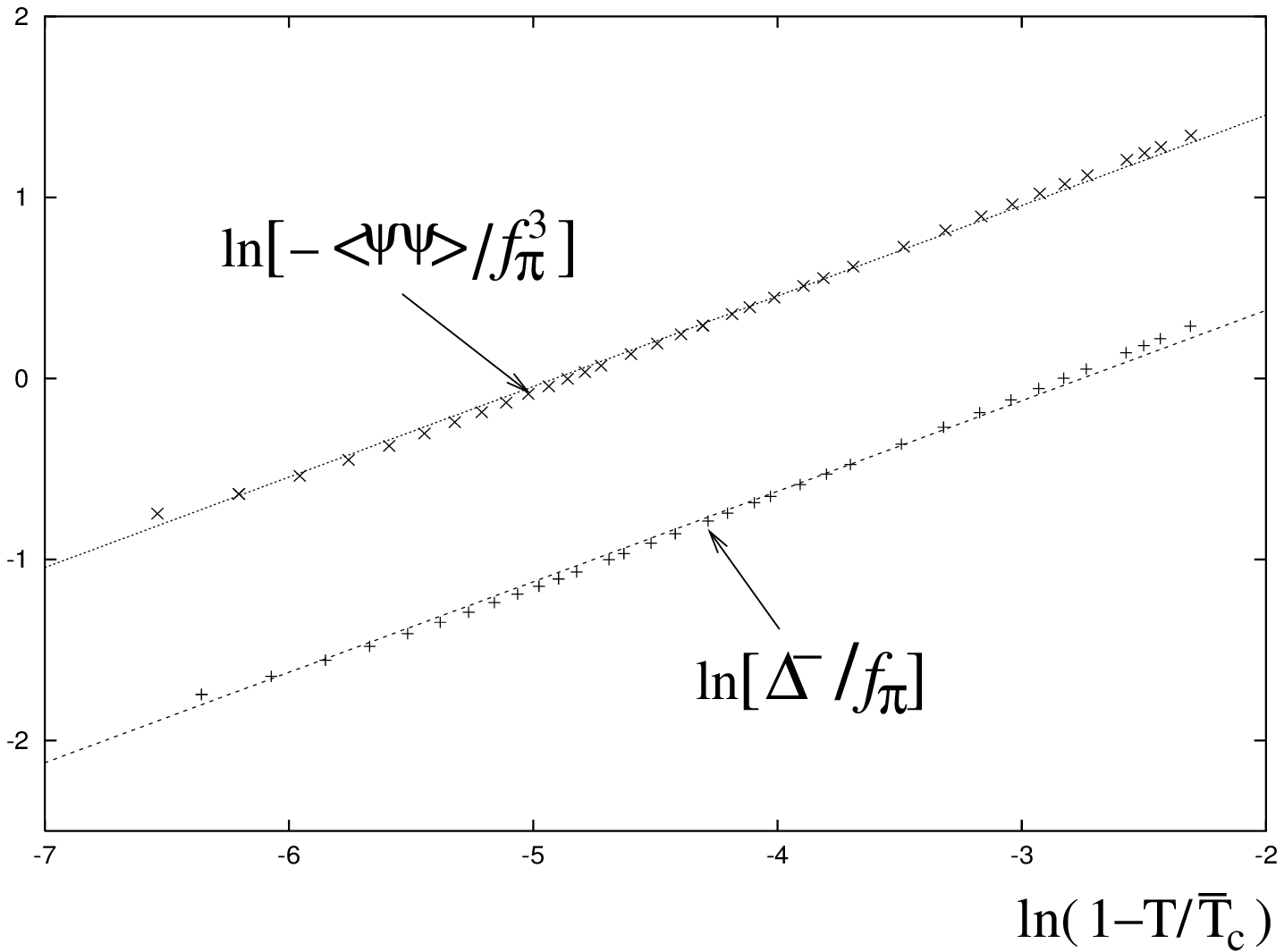}}
 \caption[]{
 Critical behavior of the condensates and the mass gaps. 
 The left panel 
 shows the critical behavior of the chiral condensate per flavor
 $-\langle\bar{\psi}\psi\rangle$ and the Dirac mass gap
 $B_1(p_0=\bar{p}=0)$ near
 the chiral second order phase transition
 at $\mu/f_\pi=0$ 
 along the arrow labeled ``chiral transition'' in 
 Fig.~\ref{phase-structure},
 while the right panel shows that of 
 the diquark condensate per flavor $-\langle\psi\psi\rangle$ and
 the Majorana mass gap $\Delta^-(p_0=0, \bar{p}=\mu)$
 near the diquark second order phase transition at $\mu/f_\pi=4$
 along the arrow labeled ``diquark transition''
 in Fig.~\ref{phase-structure}.
  The solid lines in the left panel are the
  fitting functions given in Eqs.~(\ref{meanbqq}) and (\ref{meanD}),
  and those in the right panel are the fitting functions
  given in Eqs.~(\ref{meanqq}) and (\ref{meanM}).
 }
 \label{fitDM}
\end{figure}

Let us consider the critical behavior of the condensates and the mass
gaps near the second order phase transition points
along the arrows labeled ``chiral transition'' 
($\mu=0$) and ``diquark transtion'' ($\mu/f_\pi=4$) 
shown in Fig.~\ref{phase-structure}.
First, we consider the second order chiral phase transition 
at zero chemical potential.
In Refs.~\citen{Ik,Kir1,Kir2} and \citen{Holl:1998qs}, 
it was shown, by using SDEs with different kernels,
that the scaling properties of the chiral condensate and the Dirac mass gap 
near the critical temperature are consistent 
with those obtained in the mean field treatment. 
Furthermore, the analyses done in Refs.~\citen{Tani} and \citen{Ik}
using the SDE with a similar kernel concluded that the
scaling is consistent with the mean field one,
although the numerical errors are not clear.
Thus, in the present analysis, 
we assume mean field scaling of the condensate and the mass
gap, and fit the values of 
the prefactor and the critical temperature\footnote{
We could not precisely determine the critical temperature 
through consideration of the order parameters because of the numerical error 
introduced by the discretization.}
by minimizing
\begin{equation}
 \sum_{i}|y(T_i)-F(T_i,k)|^2 \ ,
\end{equation}
where 
the index $i$ labels the temperature,
$y$ is the chiral condensate per flavor 
$-\langle\bar{\psi}\psi\rangle$
or the Dirac mass gap in the infrared limit
$B_1(p_0=\bar{p}=0)$,
and $F(T,k)$ is the fitting
function given by
\begin{equation}
F(T,k) =
k \left( 1 - \frac{T}{T_c} \right)^{1/2} \ .
\label{fitting function}
\end{equation}
Performing the fit for $0.9 {T}_c\lsim{T}<{T}_c$,
we obtain the following best fitted values for the prefactor
and the critical temperature:
\begin{eqnarray}
 &&-\langle\bar{\psi}\psi\rangle
  \sim{a}\biggl(1-\frac{T}{T_c}\biggr)^{1/2}
  \ , \quad a\sim{37.3}f_\pi^3 \ , 
  \quad {T}_c\sim{147} \ [{\rm MeV}] \ , 
  \label{meanbqq}
\\
 &&B_1(p_0=\bar{p}=0)
  \sim{\alpha}\biggl(1-\frac{T}{T_c}\biggr)^{1/2} \ ,
  \quad \alpha\sim{15.9}f_\pi \ , \quad {T}_c\sim{147} 
  \ [{\rm MeV}]  \ . 
  \label{meanD}
\end{eqnarray}
We plot the
fitting function in Eq.~(\ref{fitting function}) with the
above best fitted values in the left panel of
Fig.~\ref{fitDM}, together with the numerical data obtained
from the solution of the SDE.
This figure shows that the critical behavior of the 
chiral condensate per flavor, 
$-\langle\bar{\psi}\psi\rangle$, and the Dirac mass gap,
$B_1(p_0=\bar{p}=0)$, near $\mu=0$ 
are consistent with the mean field scaling.

We next consider the critical behavior of 
the diquark condensate per flavor $\langle\psi\psi\rangle$ and
the Majorana mass gap on the Fermi surface $\Delta^-(p_0=0, \bar{p}=\mu)$
near the color second order phase transition at
$\mu/f_\pi=4$ along the arrow labeled ``diquark transition''
 in Fig.~\ref{phase-structure}.
Using the same fitting method as used above for the
chiral transition, 
we perform the fit for $0.9 {\bar T}_c\lsim{T}<{\bar T}_c$. 
The best fitted values of the critical temperature 
and the prefactor are determined as
\begin{eqnarray}
 &&-\langle\psi\psi\rangle\sim{\bar{a}}
  \biggl(1-\frac{T}{\bar{T}_c}\biggr)^{1/2}
  \ , \quad \bar{a}\sim{11.6}f_\pi^3 
  \ , \quad \bar{T}_c\sim{102} \ 
  [{\rm MeV}] \ , 
  \label{meanqq}
\\
 &&\Delta^-(p_0=0 , \bar{p}=\mu)
  \sim{\bar\alpha}\biggl(1-\frac{T}{\bar{T}_c}
  \biggr)^{1/2} \ , \quad 
  \bar\alpha\sim{3.9}f_\pi \ , \quad \bar{T}_c\sim{102} \ 
  [{\rm MeV}] \ . 
  \label{meanM}
\end{eqnarray}
We plot the resultant fitting functions together with the 
numerical results in the right panel of Fig.~\ref{fitDM}.
This shows that the critical behavior near the
diquark phase transition point is consistent with the
mean field scaling.

\subsection{Effect of the Debye mass of the gluon}
\label{The effect of the Debye mass}

\begin{figure}[htb]
 \centerline{\includegraphics[width=12cm]{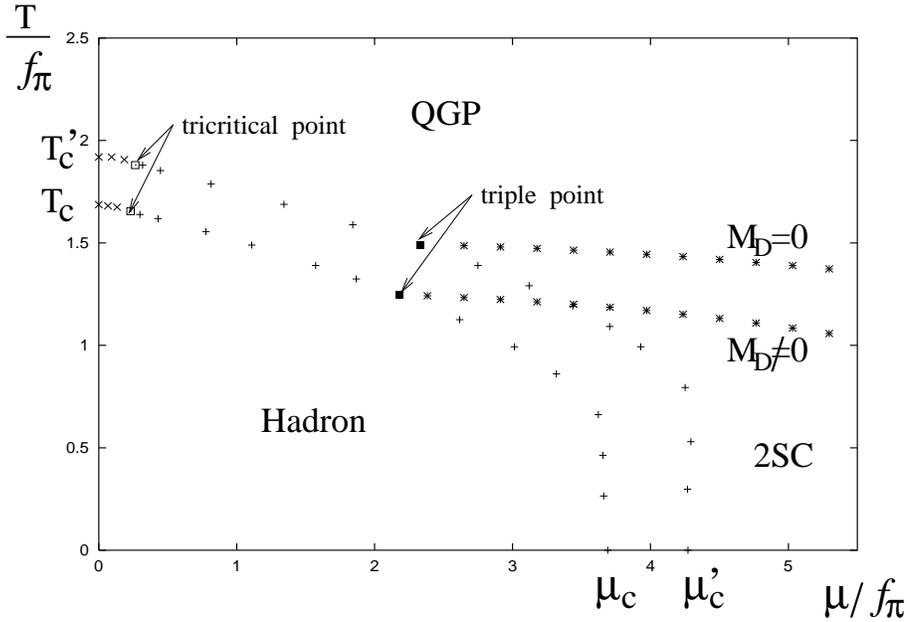}}
 \caption[]{
 Phase diagram 
 for $0\leq{T}/f_\pi\leq{2.5}$ and $0\leq\mu/f_\pi\leq{5.5}$
 obtained by setting $M_D=0$ in the SDE.
 For comparison, the phase diagram
 shown in Fig.~\ref{phase-structure} obtained with 
 the Debye mass included is also plotted.
 Here ${T}_c$=147 \mbox{MeV}, $\mu_c$=325 \mbox{MeV},
 ${T}^\prime_c$=169 \mbox{MeV}
 and $\mu^\prime_c$=376 \mbox{MeV}.}
 \label{debye-comp}
\end{figure}

In this subsection we study 
the influence of the Debye mass of the gluon on the phase diagram in case-1. 
In Fig.~\ref{debye-comp} 
we show the change of the phase diagram due to the Debye mass. 
This shows that ignoring the Debye mass does not
change the following qualitative structure 
of the phase diagram.
There is a tricritical point that separates the two branches of 
the critical line for the 
chiral phase transition from the hadron phase to the QGP phase, 
the critical line for the second order transition
and that for the first order transition.  
The phase transition from the hadron phase to the 2SC phase
is of first order, while the color phase transition 
from the 2SC phase to the
QGP phase is of second order.
However, the effect of the Debye mass changes the quantitative
structure.
In the case of zero Debye mass, 
the phase diagram without the 2SC phase
is same as that in Ref.~\citen{Ha}.
However, when we include the 2SC phase in drawing the phase diagram, 
the critical chemical potential becomes smaller than 
that in Ref.~\citen{Ha}:
At $T=0$, $\mu_c= 460\,\mbox{MeV}$ is obtained in Ref~\citen{Ha},
while we obtain $\mu_c = 376\,\mbox{MeV}$.
This is further reduced to $\mu_c = 325\,\mbox{MeV}$
by including the Debey screening mass
into the electric mode of the gluon.
Similarly, the critical line dividing
the hadron phase from the 2SC phase is moved to a region in which 
the chemical potential is smaller by about 15\% 
when the Debye mass is included. 
Furthermore, the inclusion of the Debye mass lowers the
value of the critical temperature at the phase transition from
the hadron phase to the QGP phase by about 15\%,
while it lowers the value of the critical temperature 
at the phase transition from the 2SC phase to the QGP phase 
by 15$-$30\%.
Note, however, that the value of the chemical potential at
the tricritical point is not changed, although the value of 
the critical temperature at this point decreases by about 15\%.
These quantitative changes imply 
that the Debye mass of the electric mode 
plays a role to weaken the attractive interaction between two quarks
as well as that between a quark and an antiquark.
This is reasonable, because the Debye mass screens 
the long range force mediated by the electric mode of a gluon. 
Therefore we expect 
that the critical temperatures of the chiral and color
phase transition from the hadron phase and the 2SC phase 
to the QGP phase may be further reduced
if the contact interaction, i.e., the short-range force
induced by the instanton, is the dominant one in this
intermediate chemical potential region.

\subsection{Effect of the antiquark contribution}
\label{antiquark contribution}

In this subsection we study 
the effect of the antiquark contribution by solving
the following three different types of SDEs:
(case-1) Coupled SDEs 
 for $\Delta^-$ and $\Delta^+$;
(case-2) the SDE for $\Delta^-$ with $\Delta^+=0$;
(case-3) the SDE for $\Delta^-$ with $\Delta^+=\Delta^-$.
In case-1, the quark mass and the antiquark mass
are included properly, while the antiquark mass is ignored in case-2. 
The approximation in case-2 is considered 
to be valid in the high density region.
The approximation in case-3 is used in
many analyses carried out using models with the local 4-Fermi interaction. 

\begin{figure}[htb]
 \centerline{\includegraphics[width=12cm]{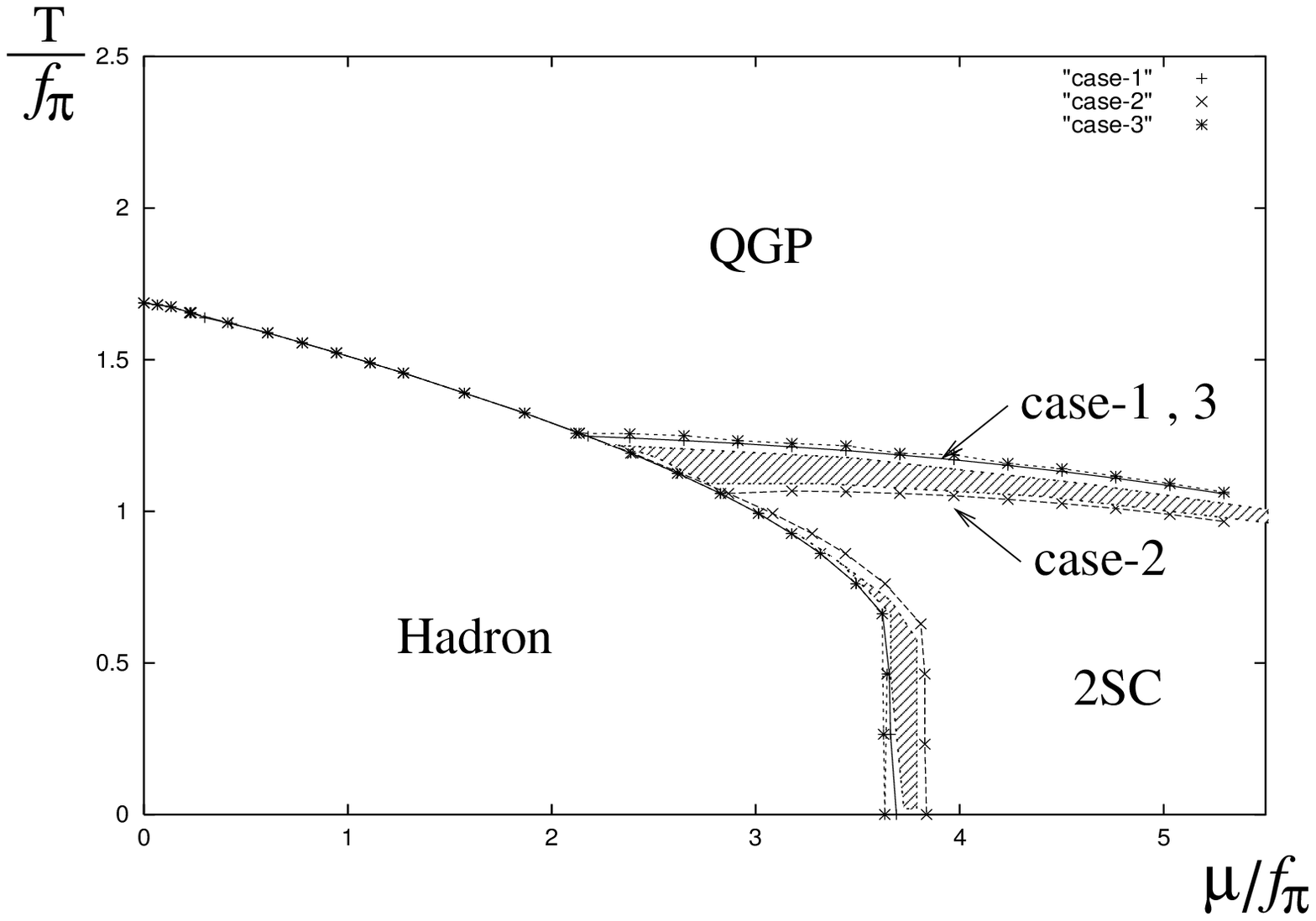}}
 \caption{Phase diagrams in three different cases 
 for $0\leq{T}/f_\pi\leq{2.5}$ and $0\leq\mu/f_\pi\leq{5.5}$. 
 The symbol $+$ denotes the phase transition in case-1,
 $\times$ that in case-2, and $\times\hspace{-0.75em}+$ that in case-3.
 The points in case-1, case-2 and case-3 are connected by 
 solid curves, dashed curves and dotted curves, respectively. 
 The shaded areas are in the 2SC phase in case-1 
 but not in case-2.
}
 \label{pha-3case2}
\end{figure}
We show the phase diagrams of above three different cases 
in Fig.~\ref{pha-3case2}. 
In this figure, the symbol $+$ denotes 
the phase transtion in case-1,
$\times$ that in case-2 and $\times\hspace{-0.75em}+$ 
that in case-3. To make the differences among the three cases clear,
we connect the data points by a solid curve in case-1,
by a dashed curve in case-2, and by a dotted curve in case-3. 
First of all, we should note that
the critical line between the hadron phase
and the QGP phase for $\mu/f_\pi\lsim 2$ is not at all affected 
by the change of the antiquark contribution, because
there exist no 2SC solutions.
Therefore, in all three cases, we obtained the same tricrical point
at $(T,\mu)=(146,20)\,\mbox{MeV}$, where the
critical line is divided into that for the second order phase
transition ($\mu<20\,\mbox{MeV}$)
and that for the first order phase transition
($20\,\mbox{MeV}<\mu\lsim 180\,\mbox{MeV}$).
Similarly to the effect of Debey mass studied in the previous subsection,
the effect of the antiquark contribution does not change 
the following qualitative structures of the phase diagram:
In all three cases,
the phase transition from the hadron phase to the 2SC phase is of
first order,
while the color phase transition from the 2SC phase
to the QGP phase is of second order.
Furthermore,
the approximation represented by setting $\Delta^+=\Delta^-$ (case-3)
results in almost the same phase diagram as in the case of 
the full analysis (case-1),
except that the critical chemical potential at $T\simeq0$ in case-3
is slightly smaller than that in case-1.
However, the omission of the
antiquark contribution (case-2)
quantitatively
changes the critical line between the hadron phase and the 2SC phase,
as well as that between the 2SC phase and the QGP phase.
In Fig.~\ref{pha-3case2}, we indicate the region 
in the 2SC phase in case-1 
(but not in case-2) by the shaded areas. 
This clearly shows that the 
region of the 2SC phase in case-2 is smaller than that in case-1, 
as well as in case-3. 
The value of the critical temperature at the color phase transtion in
case-1 becomes larger by about 10\% at $\mu/f_\pi=4$ than that in case-2. 
The value of the critical chemical potential at the
chiral phase transition in case-1 becomes smaller by about 5\%
at zero temperature than that in case-2.

Let us study the Majorana mass gap and the diquark condensate
at zero temperature.
In the remainder of this subsection, we study them
not only in the true vacuum (i.e., in the case that
the 2SC vacuum is the most stable)
but also in the false vacuum (i.e., in the case that the 2SC vacuum is
less stable than the $\chi$SB vacuum)
in order to see the effect of the antiquark contribution more clearly.

\begin{figure}[htb]
 \centerline{
 \includegraphics[width=7cm]{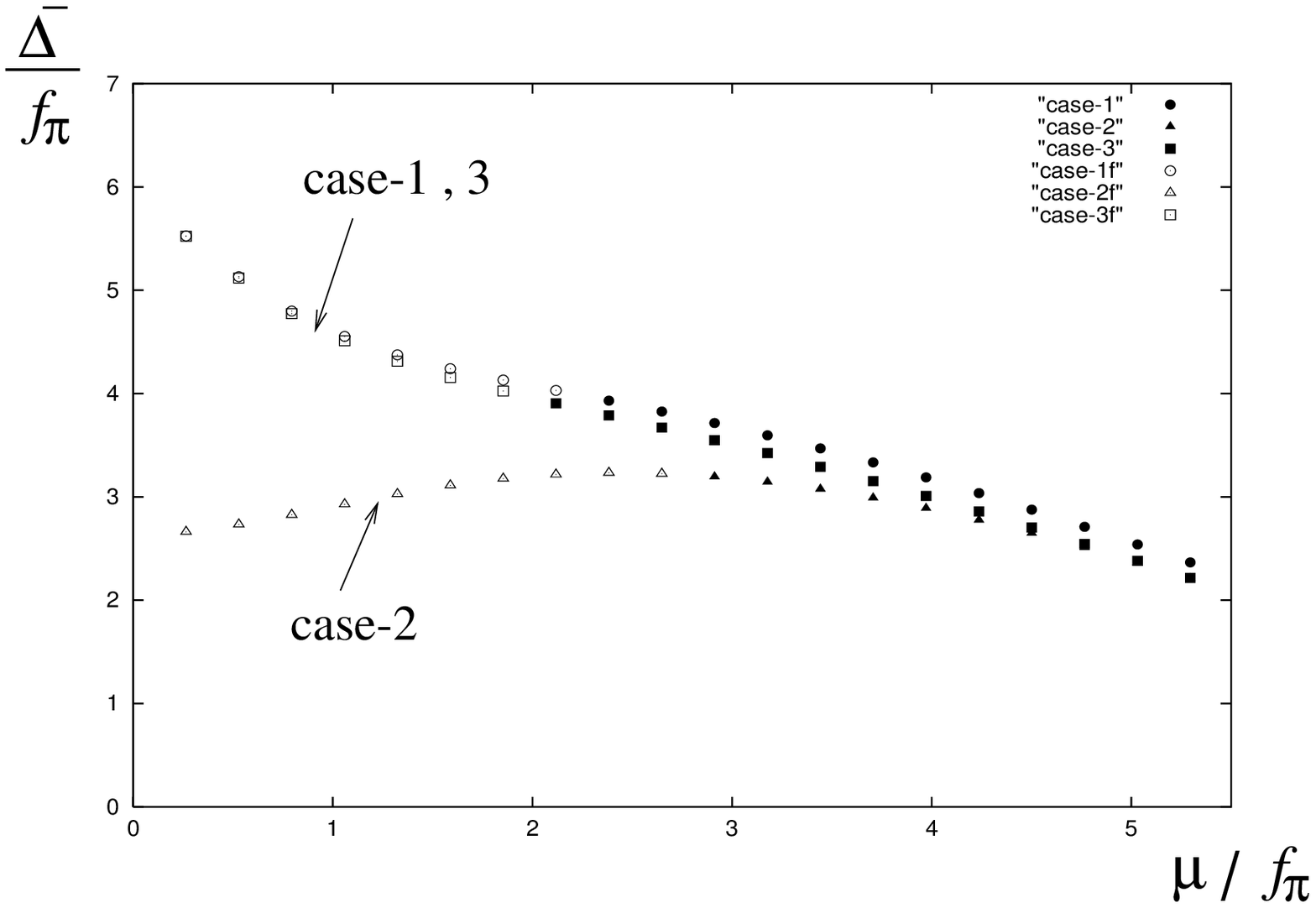}
 \includegraphics[width=7cm]{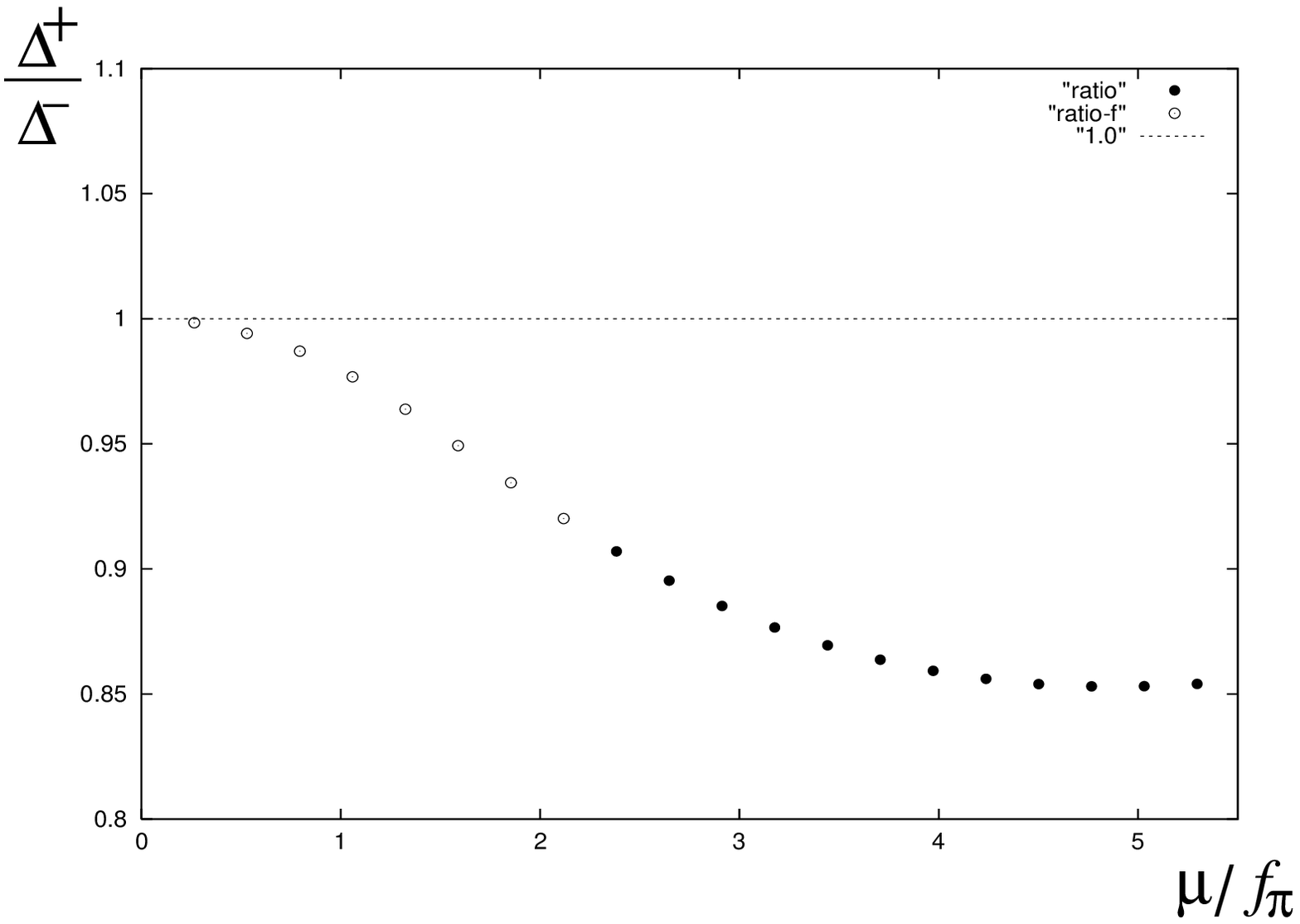}
 }
 \caption{
 Dependence on the chemical potential of
 $\Delta^-(p_0=0,\bar{p}=\mu)$ in three cases (left panel)
  and $\Delta^+(p_0=0,\bar{p}=\mu)$/$\Delta^-(p_0=0,\bar{p}=\mu)$
 in case-1 (right panel)
 for $0\leq\mu/f_\pi\leq{5.5}$ 
 and $T=0$.
  In the left panel,
  the data points indicated by the circles, triangles and squares
  were obtained in case-1, case-2 and case-3, respectively.
  Open symbols ({\footnotesize $\bigcirc$}, 
  {\small $\triangle$} and {\small $\square$}) represent
  data points in the false vacuum 
  (the case in which the 2SC vacuum is less stable than the $\chi$SB vacuum),
  while the filled ones represent data points in the true vacuum
  (the case in which the 2SC vacuum is most stable).
 }
 \label{majorana3}
\end{figure}
In Fig.~\ref{majorana3} 
we show the dependence on the chemical potential of the quark 
Majorana mass gap $\Delta^-$ on the Fermi surface (left panel) and 
the ratio of the antiquark mass to the quark mass 
$\Delta^+/\Delta^-$ at $p_0=0$ and $\bar{p}=\mu$
in case-1 (right panel) at zero temperature. 
We note that a nontrivial solution 
exists in all cases, even at $\mu=0$, because
the running coupling in the infrared region
exceeds the critical value, $\pi/2$,
as discussed in Subsection~\ref{Phase Structure}.
Furthermore,
the ratio $\Delta^+/\Delta^-$ is actually 1 at $\mu=0$,
which is required by the existence of charge conjugation symmetry.
As a result, $\Delta^-$ in case-3 is equal to 
that in case-1 at $\mu=0$.
However, the value of $\Delta^-$ in case-2 is about half of that in
case-1 and case-3 at $\mu=0$.
When we increase the chemical potential,
the values of $\Delta^-$ in case-1 and case-3 decrease,
while that in case-2 first increases and then decreases.
The right panel of Fig.~\ref{majorana3} shows that the
ratio $\Delta^+/\Delta^-$ in case-1
decreases as $\mu$ increases and reaches a value of about $0.85$.
The left panel of Fig.~\ref{majorana3} shows that,
for $\mu/f_\pi \gsim 3.5$,
the values of $\Delta^-$ in all three cases are almost same,
although the values of $\Delta^+$ differ greatly.
In fact, we have $\Delta^+(\mbox{case-1}):\Delta^+(\mbox{case-2}):
\Delta^+(\mbox{case-3}) \simeq 0.85:0:1$.
These results imply that
the antiquark gives a sizable contribution
to the quark Majorana mass
in the low chemical potential region, 
while it becomes negligible for $\mu/f_\pi \gsim 3.5$.
This suggests that,
to form the Majorana mass in the
region where the 2SC vacuum is most stable,
the effect of the Fermi surface
is more important than the effect of the large attractive force
due to the large running coupling.

\begin{figure}[htb]
 \centerline{\includegraphics[width=12cm]{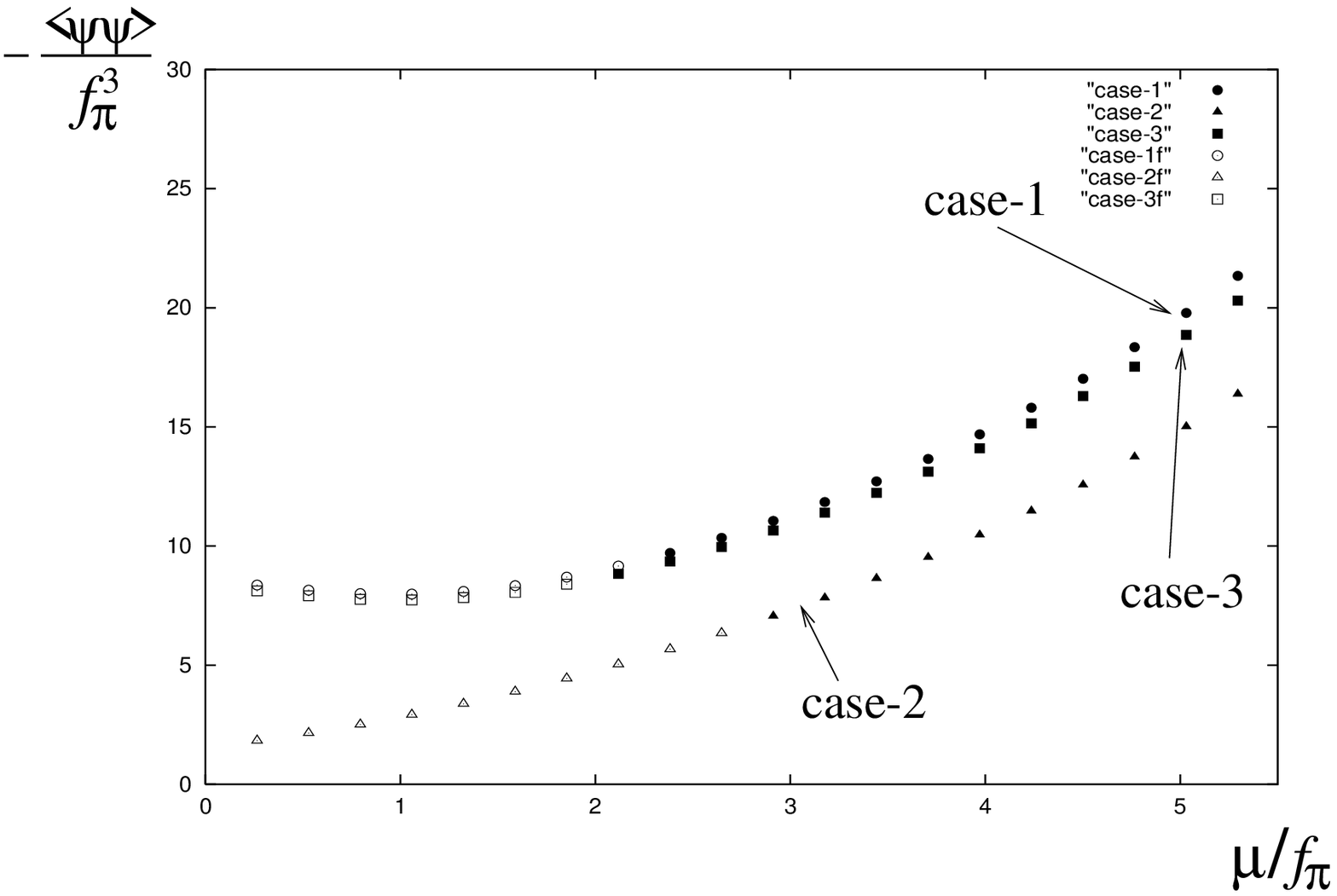}}
 \caption{
  Dependence on the chemical potential of the diquark condensates
  per flavor for $0\leq\mu/f_\pi\leq{5.5}$ 
  and $T=0$.
  The data points indicated by the circles, triangles and squares
  were obtained in case-1, case-2 and case-3, respectively.
  The open symbols ({\footnotesize $\bigcirc$}, 
  {\small $\triangle$} and {\small $\square$}) represent
  data points in the false vacuum 
  (the case in which the 2SC vacuum is less stable than the $\chi$SB vacuum),
  while the filled ones represent data points in the true vacuum
  (the case in which the 2SC vacuum is most stable).
 }
 \label{ratio-deltanp}
\end{figure}
Next, we show the dependence on the chemical potential 
of the diquark condensates per flavor in three cases
in Fig.~\ref{ratio-deltanp}.
This shows that
the value of the diquark condensate in case-3 ($\Delta^+=\Delta^-$)
is almost same as that in case-1,
which is consistent with the fact that in case-1
the value of $\Delta^+$ is comparable to that of $\Delta^-$.
However, there is a large difference between 
the value of the diquark condensate in case-2 and that obtained
in case-1 or case-3.
Actually, the ratio of the value of the diquark condensate in case-1 or case-3
to that in case-2 is roughly 4
in the region of very small chemical potential:
$\left\langle\psi\psi\right\rangle_{\mbox{\scriptsize case-1}}/
\left\langle\psi\psi\right\rangle_{\mbox{\scriptsize case-2}}
\simeq 4$ for $\mu \simeq 0$.
This 
can be understood as follows.
Figure~\ref{majorana3} shows that
the value of $\Delta^-$ in case-1 is about twice that
in case-2 for $\mu \simeq 0$, 
and that $\Delta^+$ is comparable to $\Delta^-$
in case-1 for $\mu \simeq 0$.
On the other hand, 
the formula for calculating the diquark condensate given
in Eq.~(\ref{diquark formula}) shows that the diquark condensate
consists of quark and antiquark contributions, and that
the dominant part comes from the ultraviolet region.
Thus, the quark and antiquark contributions to the diquark
condensate in case-1 are each about twice the quark contribution
in case-2, 
and therefore the ratio of the value of the diquark condensate in case-1
or case-3 to that in case-2 is about 4 near $\mu=0$.
Now, how about the ratio
in the region of intermediate chemical potential ($\mu/f_\pi \gsim 3.5$)?
If the effect of the Fermi surface were negligible,
the same argument as above would lead us to conclude that
the values of diquark condensate in case-3 and case-1
are roughly twice that in case-2.
However, Fig.~\ref{ratio-deltanp} shows that 
the ratio is actually smaller than 2, and it becomes
about $4/3$ at $\mu/f_\pi =5$: 
$\left\langle\psi\psi\right\rangle_{\mbox{\scriptsize case-1}}/
\left\langle\psi\psi\right\rangle_{\mbox{\scriptsize case-2}}
\simeq 4/3$ for $\mu/f_\pi=5$.
This implies that, as expected, the effect of the
Fermi surface plays an important role in the region of intermediate 
chemical potential.

\begin{figure}[htb]
 \centerline{
 \includegraphics[width=12cm]{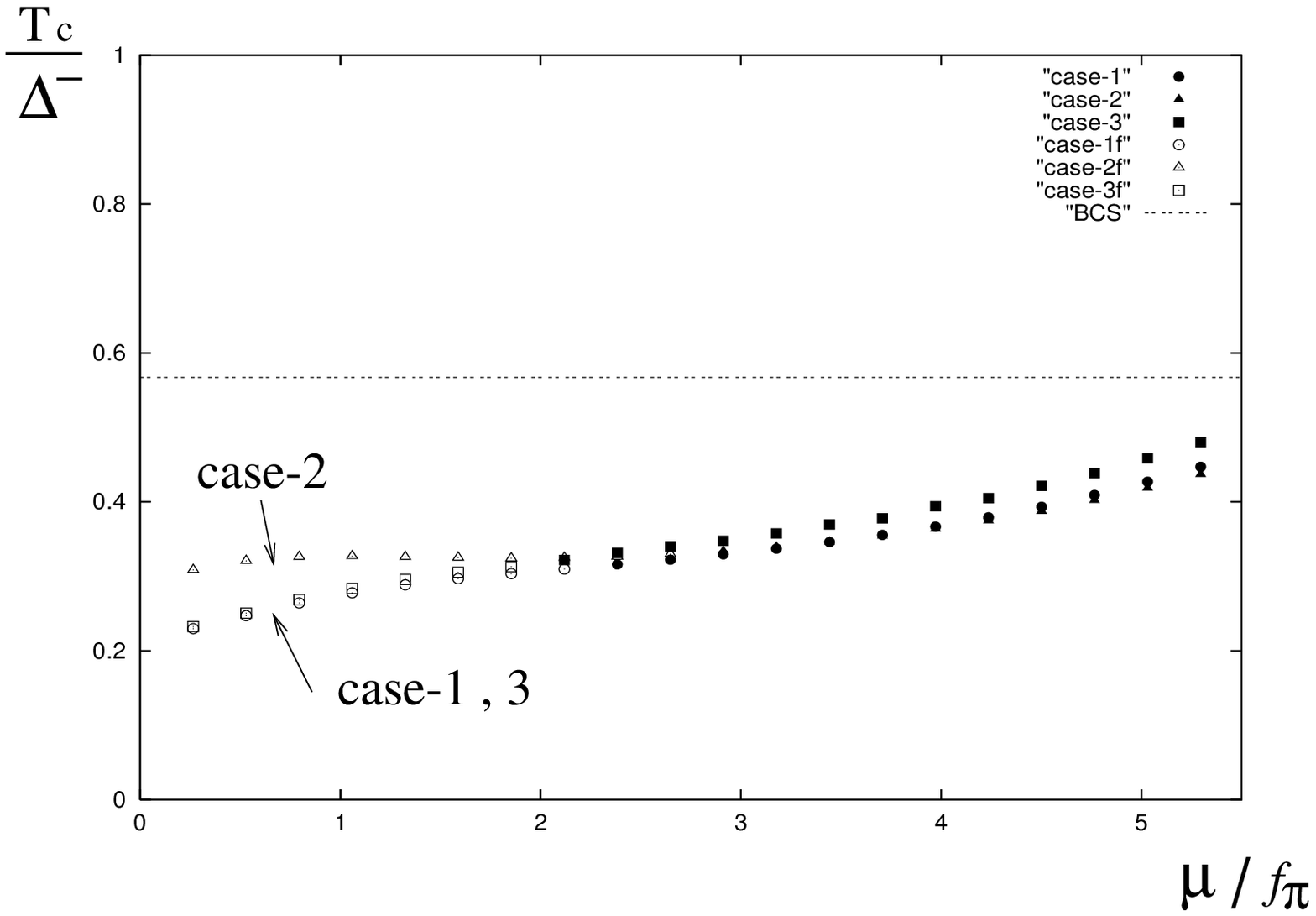}
 }
 \caption[]{
 Dependence on the chemical potential of the ratios of $T_c$ to 
 $\left.\Delta^-(p_4=0;\bar{p}=\mu)\right\vert_{T=0}$
 for $0\leq\mu/f_\pi\leq{5.5}$. 
 The data points indicated by the circles, triangles and squares
 were obtained in case-1, case-2 and case-3, respectively.
 The open symbols ({\footnotesize $\bigcirc$}, 
 {\small $\triangle$} and {\small $\square$}) represent 
 data points in the false vacuum 
 (the case in which the 2SC vacuum is less stable than the $\chi$SB vacuum),
 while the filled ones represent data points in the true vacuum
 (the case in which the 2SC vacuum is most stable).
 }
 \label{Tc-over-Delta3}
\end{figure}

The results we have found in this subsection to this point
(Figs.~\ref{pha-3case2}$-$\ref{ratio-deltanp}) imply that
the antiquark gives a non-negligible contribution,
and that the approximation $\Delta^-=\Delta^+$ is sufficiently valid 
to allow a study of the phase diagram, the quark Majorana mass gap 
and the diquark condensate
in the region of intermediate temperature and intermediate chemical
potential, where the chiral phase transition occurs.
The same approximation may be sufficiently valid
to allow investigations of other physical quantities
such as the number density, as well.

Let us compare the results in the ratio of the critical temperature to 
the Majorana mass gap at zero temperature with the BCS result, 
${T}_c/\Delta \simeq 0.567$,
obtained in the high density region~\cite{pis,Bailin:bm}.
In Fig.~\ref{Tc-over-Delta3} we show the 
dependence on the chemical potential of 
the ratios of the critical temperature 
to the quark Majorana mass gap on the Fermi surface
at zero temperature. 
There are only small differences among the data points obtained
in the three cases.
{}From this figure, we see that 
this ratio approaches the BCS 
value, $0.567$, when we increase the value of the chemical potential,
and that it is already close to the BCS value
in the intermediate chemical potential region:
\begin{equation}
{T}_c/\Delta^- \sim 0.5 \ ,
\quad \mbox{for} \ \mu\sim 450\,\mbox{MeV} \ .
\end{equation}

\begin{figure}[htb]
 \centerline{
 \includegraphics[width=7cm]{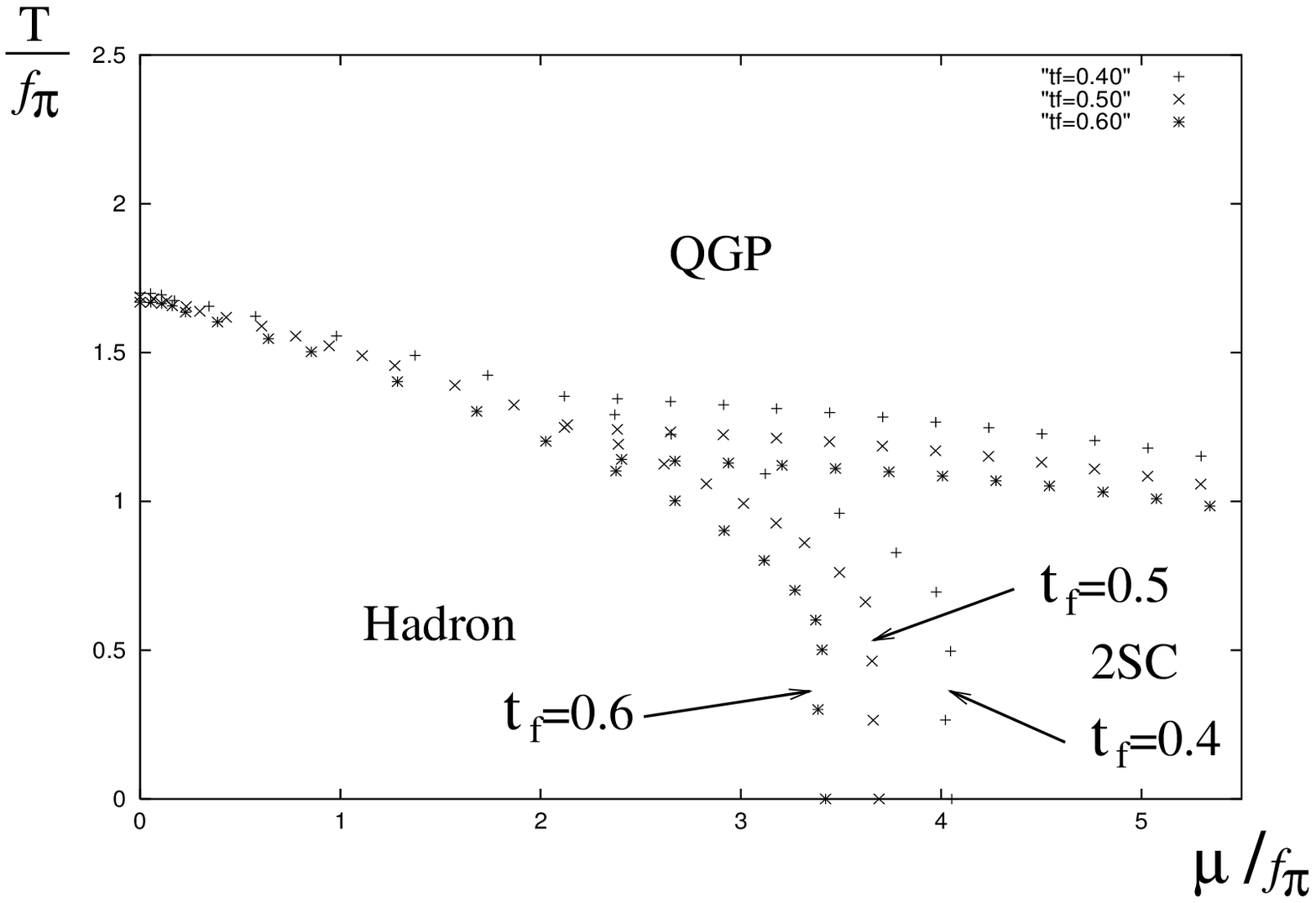}
 \includegraphics[width=7cm]{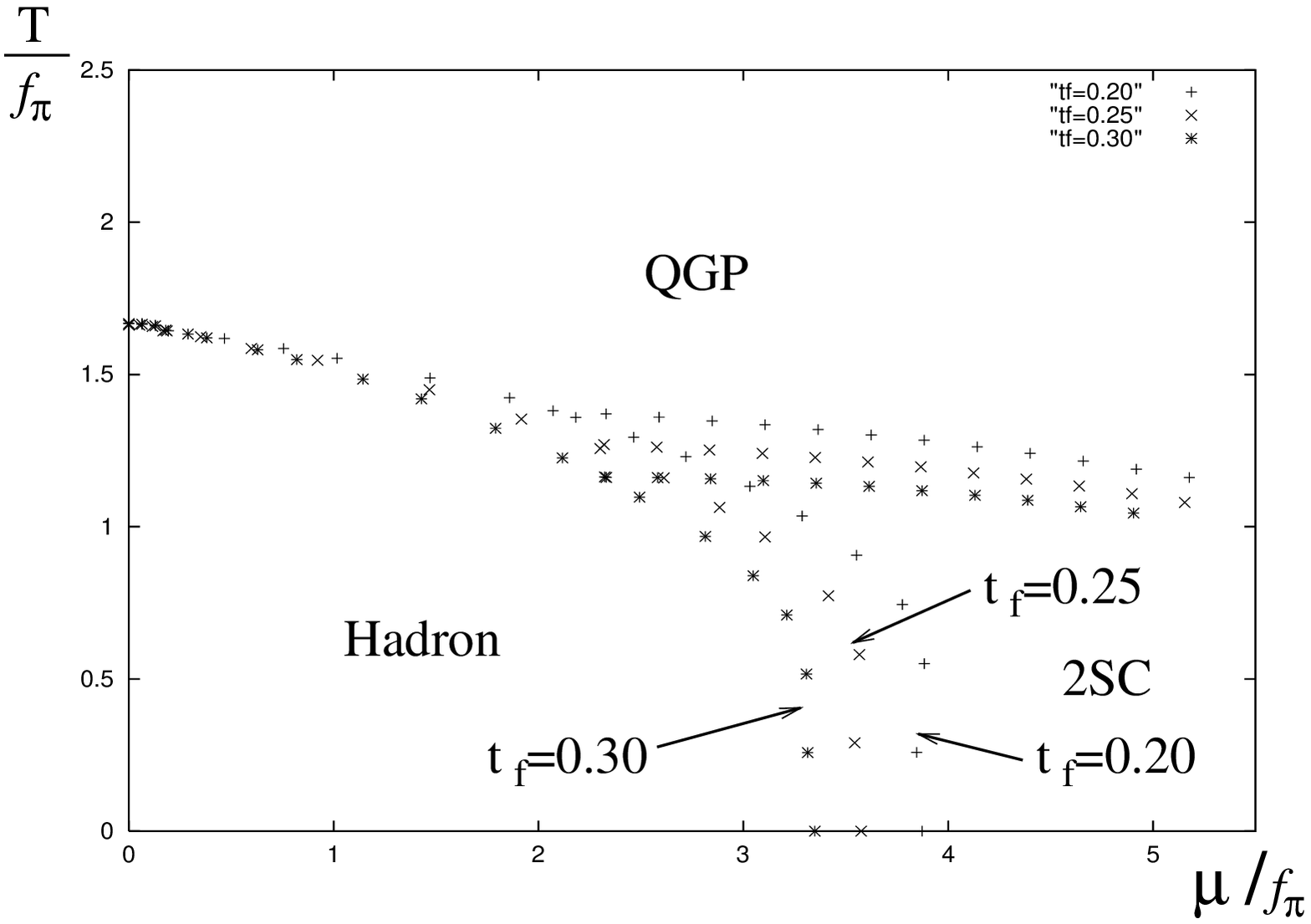}
 }
 \caption{Phase diagram for several choices of
 the infrared cutoff parameter $t_f$. 
 The left panel plots the critical lines for $t_f=0.4$, $0.5$ 
 and $0.6$ with the running coupling of type (I), given in Eq.~(\ref{run}).
 The right panel plots the critical lines for $t_f = 0.2$, $0.25$ and
 $0.3$ with the running coupling of type (II), given in Eq.~(\ref{run2}).
}
 \label{tf-depen}
\end{figure}

Finally, we check the dependence of the phase diagram
on the model used by changing the infrared regularization parameters for
two types of running couplings given in Eqs.~(\ref{run}) and (\ref{run2}). 
We show the phase diagrams 
for the running coupling of type (I) 
with $t_f=0.4$, $0.5$ and $0.6$ in the left panel of 
Fig.~\ref{tf-depen} 
and the phase diagrams 
for the running coupling of type (II) 
with $t_f=0.20$, $0.25$ and $0.30$ in the right panel of 
Fig.~\ref{tf-depen}.
These figures show that
the critical line for the phase transition from the hadron phase
to the 2SC phase, as well as that from the 2SC phase to the
QGP phase, has a weak dependence on the infrared regularization
parameter $t_f$:
For a larger value of $t_f$, the phase transition from the 
hadron phase to the 2SC phase occurs at a smaller chemical
potential, and that from the 2SC phase to the QGP phase
occurs at a smaller temperature.
Contrastingly, the critical line for the 
chiral phase transition from the hadron phase to the QGP phase
is very stable against changes in $t_f$
for both types of the running coupling.
Furthermore, the running coupling of type (II) 
with $t_f = 0.25$
gives almost the
same critical line for the chiral phase transition
as that of type (I) with $t_f=0.5$.
These results imply that the phase structure in the region of 
small chemical potential is very insensitive
to the infrared regularization of the running coupling.

\section{Summary and discussion}
\label{Summary and Discussion}

We studied the phase structure of hot and/or dense QCD by solving the 
Schwinger-Dyson equations for the Dirac and Majorana masses
of the quark propagator with the improved ladder approximation in the
Landau gauge.
We found that there exists a tricritical point at 
$({T}\,,\, \mu)=(146, 20)\,\mbox{MeV}$, which divides
the critical line between the hadron phase and the QGP phase
into the line of a second order phase transition 
[between $({T}\,,\, \mu)=(147, 0)\,\mbox{MeV}$
and $(146, 20)\,\mbox{MeV}$]
and that of a first order phase transition 
($20\,\mbox{MeV}< \mu $).
Our result implies that
at the second order phase transition at $\mu=0$,
the scaling properties of 
the Dirac mass and the chiral condensate are consistent
with the mean field scalings.

The phase transition from the hadron phase to the 2SC phase
was found to be of first order at non-zero temprature,
as we have obtained at zero temperature
in previous analysis~\cite{Taka}.
Furthermore, we found that
the color phase transtion from the 2SC phase to the
QGP phase is of second order, with the scaling properties of
the Majorana mass and the diquark condensate consistent with
the mean field scalings.
The resultant critical temperature of the color phase
transition is
on the order of $100\,\mbox{MeV}$,
which is about twice the value obtained in 
analyses carried out with models based on the contact 4-Fermi 
interaction (see, e.g., Refs.~\citen{Be,Sc,Kit,Kit2}).
We believe that this increase of the critical temperature may be 
caused by the long range force mediated by the magnetic
mode of the gluon.

In the present paper we performed analysis 
with the imaginary part of the Dirac mass included, 
because in the SDE at non-zero chemical potential, 
the imaginary part, which is momentum dependent 
(an odd function in $p_0$), is inevitably 
generated in the hadron ($\chi{SB}$) phase.
However, some analyses of the SDE do not include   
the imaginary part, and analyses done using the local 4-Fermi 
interaction model do not generally include the imaginary part, 
because a leading order approximation is used.
We found that the most noteworthy feature of the analysis that includes
the imaginary part of the Dirac mass regards 
the position of the tricritical point in the $T$-$\mu$ plane.
When we use the SDE without the imaginary
part of the Dirac mass (i.e., ${\rm Im}[B(p)]=0$),
the tricritical point is at $({T}, \mu)=(124, 210) \mbox{MeV}$
[and the triple point is at $({T}, \mu)=(111, 243) \mbox{MeV}$].
The value of $\mu$ at this tricritical point 
is close to the values [$\mu\sim{O(100) \mbox{MeV}}$] 
obtained from analyses carried out using models 
with the local 4-Fermi interaction~\cite{Be,Sc,Kit,Kit2,Va} 
and those carried out using the SDEs with a momentum dependence 
of the mass function assumed~\cite{Bar,Kir1}. The result here implies that,
in the SDE analysis, including the imaginary part 
causes the tricritical point to move to a position of smaller 
chemical potential: $({T}, \mu)=(124, 210) \mbox{MeV} 
\rightarrow ({T}, \mu)=(146, 20) \mbox{MeV}$.
Therefore, we believe that the imaginary part of the Dirac mass is important 
and should be included in the SDE at finite chemical potential.

We studied the effect of the Debye mass of the gluon, 
and showed how the phase structure is changed.  
When we ignore the Debye mass, the critical temperature 
for the color phase transition 
from the 2SC phase to the QGP phase 
is increased by about 
15$-$30\%.
In addition, the critical temperature ${T}_c$ 
for the chiral phase transition 
from the hadron phase to the QGP phase and 
the critical chemical potential $\mu_c$
for the phase transition from the hadron phase to the 2SC phase
are also increased by about 15\%.

We examined the effect of antiquark contribution.
Our results show that 
the antiquark Majorana mass gap, $\Delta^+$,
is comparable to the quark one, $\Delta^-$,
in all the chemical potential regions that we studied:
$1 > \Delta^+/\Delta^- \gsim 0.85$ for 
$0 < \mu \lsim 500\,\mbox{MeV}$.
As a result, the approximation represented by ignoring the antiquark mass,
which is generally considered to be good at very high density,
results in quantitative differences in the phase diagram,
the value of the Majorana mass gap, and that of the diquark condensate.

By contrast, analysis employing the approximation 
represented by setting $\Delta^+=\Delta^-$,
which is often adopted in analysis using models with
the local 4-Fermi interaction,
gives almost the same results
for the phase diagram, the quark Majorana mass gap and the diquark 
condensate as those obtained using the full analysis, 
in spite of the fact that the value of $\Delta^+$ is about 15\% 
smaller than that of $\Delta^-$ in the full analysis.
Presumably, in regions of intermediate temperature 
and intermediate chemical potential,
it is also sufficient to set the antiquark Majorana 
mass equal to the quark one $(\Delta^+=\Delta^-)$
for the purpose of studying other physical quantities,
such as the number density.

As in previous analysis done at ${T}=0$~\cite{Taka},
we sought the mixed phase, in which both the chiral condensate and
the diquark condensate exist, 
by solving the coupled SDEs for Majorana
and Dirac masses, starting from several initial trial functions.
However, we could not find such a solution in the present analysis.

It is important to investigate the QCD phase structure 
by studying phenomena in compact stars and heavy ion collisions.
In this paper, we find the color phase transition to be of second order.
If this is the case, there is a stronger possibility for 
the existence of a pseudogap phase,
as discussed in Ref.~\citen{Kit}.

In this paper we ignored the
Landau damping of the magnetic mode of the gluon,
because the approximated form adopted in the previous
analysis~\cite{Taka} may not be valid in the region of 
low chemical potential.
It may be interesting to study the effect of Landau damping
by including it as a
hard thermal and/or dense loop correction without further
approximation,
although we may have to perform the angular integration of the SDE
numerically in such a case.

We assumed that
the Landau gauge causes no deviation of
the wave function renormalization of the quark propagator from 1,
even at non-zero temperature and/or non-zero chemical potential.
In order to keep the wave function renormalization equal to 1,
we may have to introduce
a non-local gauge fixing term, as in Ref.~\citen{Kugo:1992pr},
at $T=\mu=0$.
We expect that this will not change the qualitative structure of 
the present results.
In this paper, we have the QCD scale 
$\Lambda_{\rm qcd}\sim{600}$ ${\rm MeV}$,
which is larger than the value 
determined from the experimental value of $\alpha_s$ 
in the high energy region,
although the value of the pion decay constant $f_\pi$ is 
set to a value consistent with experiments.
In Ref.~\citen{Hashimoto:2002px}, 
an effective coupling that includes higher order
corrections is used to show that the values of $\Lambda_{\rm qcd}$ and $f_\pi$
become consistent with experiments.
It would be interesting to use such a running coupling to analyze 
hot and/or dense QCD.

\section*{Acknowledgements}

The author is very grateful to M. Harada for helpful discussions
and careful reading of this manuscript.
The author thanks M. Alford for a useful comment.

\appendix
\section{Condensates, Effective Potential and Schwinger-Dyson Equation}
\label{Quark Propagator and the Schwinger-Dyson Equation}

In this appendix we give explicit forms of condensates, the effective
potential and SDEs. The integration kernels in the SDEs are different from
those in Ref.~\citen{Taka}, because the forms of the gluon propagators
used in the SDEs are different. 

The explicit forms of the chiral condensate and the diquark condensate
are given by
\begin{eqnarray}
 &&\langle{\Omega}\vert\bar\psi_{a}^i\psi_{i}^a(0)
  \vert{\Omega}\rangle_{\Lambda} \nonumber\\
 &&\hspace{1cm}=4N_f
  T\sum_{n}\int^\Lambda\frac{d^3p}{(2\pi)^3} \nonumber\\
 &&\hspace{1.5cm}\biggl[\frac{N_c-1}{F(p,B_1,\Delta)}
  \biggl\{\biggl((p_0-\mu)^2-\bar{p}^2-\{B_1(-p)\}^2\biggr)B_1(p)
  -\Delta^+(p)\Delta^-(p)B_1(-p)\biggr\} \nonumber\\
  &&\hspace{2cm}+\frac{1}{F(p,B_3,\Delta=0)}
  \biggl((p_0-\mu)^2-\bar{p}^2-\{B_3(-p)\}^2\biggr)B_3(p)
  \biggr] \ , 
\label{app:chiral condensate}
\\
 &&\langle{\Omega}\vert(\epsilon^{ij}\epsilon_{ab3})
  [\psi^{T}]^a_iC\gamma_5\psi_j^b(0)\vert{\Omega}\rangle_{\Lambda} \nonumber\\
 &&\hspace{1cm}=4(N_c-1)N_f{T}\sum_{n}\int^\Lambda\frac{d^3p}{(2\pi)^3}
  \frac{1}{F(p,B_1,\Delta)}\frac{1}{2} \nonumber\\
 &&\hspace{3cm}
  \biggl[\biggl\{(p_0)^2-(\bar{p}+\mu)^2-\{\Delta^+(p)\}^2
  -\vert{B_1}(p)\vert^2\biggr\}\Delta^-(p) \nonumber\\
 &&\hspace{3cm}+\biggl\{(p_0)^2-(\bar{p}-\mu)^2-\{\Delta^-(p)\}^2
  -\vert{B_1}(p)\vert^2\biggr\}\Delta^+(p)\biggr] \ ,
\label{app:diquark condensate}
\end{eqnarray}
where $F$ is defined as
\begin{eqnarray}
 F(p,B,\Delta)
  &=&[(p_0+\mu)^2-\bar{p}^2-\{B(p)\}^2]
  [(p_0-\mu)^2-\bar{p}^2-\{B(-p)\}^2] \nonumber\\
 &&\hspace{0.5cm}-[(p_0)^2-(\bar{p}-\mu)^2]\vert\Delta^+(p)\vert^2
  -[(p_0)^2-(\bar{p}+\mu)^2]\vert\Delta^-(p)\vert^2 \nonumber\\
  &&\hspace{1cm}+\vert\Delta^+(p)\vert^2\vert\Delta^-(p)\vert^2
  +2B(p)B(-p)\Delta^+(p)\Delta^-(p) \ . 
\end{eqnarray}
The chiral condensate in Eq.~(\ref{app:chiral condensate})
takes the following form
 when we set $\Delta=0$ and $B_1=B_3=B$:
\begin{eqnarray}
 \langle{\Omega}\vert\bar\psi_{a}^i\psi_{i}^a(0)
  \vert{\Omega}\rangle_{\Lambda}
  &=&\langle{\Omega}\vert[\bar\psi_C]_{a}^i[\psi_C]_{i}^a(0)
  \vert{\Omega}\rangle_{\Lambda} \nonumber\\
 &=&4N_cN_f{T}\sum_{n}\int^\Lambda\frac{d^3p}{(2\pi)^3}
  \frac{B(p)}{(p_0+\mu)^2-\bar{p}^2-\{B(p)\}^2} \ , 
\end{eqnarray}
where $N_c=3$ and $N_f=2$.
The diquark condensate in Eq.~(\ref{app:diquark condensate})
takes the following form
 when we set $B_1=B_3=0$:
\begin{eqnarray}
 &&\hspace{1cm}\langle{\Omega}\vert
  [\psi^T]_i^aC\gamma_5\psi_j^b(0)\vert{\Omega}\rangle_{\Lambda}
  =-\langle{\Omega}\vert(\epsilon^{ij}\epsilon_{ab3})
  [\psi_C^T]_i^aC\gamma_5[\psi_C]_j^b(0)\vert{\Omega}\rangle_{\Lambda} 
  \nonumber\\
 &&\hspace{1.5cm}=4(N_c-1)N_f{T}\sum_{n}\int^\Lambda\frac{d^3p}{(2\pi)^3}
  \nonumber\\
 &&\hspace{2cm}
  \frac{1}{2}
  \biggl[\frac{\Delta^-(p)}{(p_0)^2-(\bar{p}-\mu)^2-\{\Delta^-(p)\}^2}
  +\frac{\Delta^+(p)}{(p_0)^2-(\bar{p}+\mu)^2-\{\Delta^+(p)\}^2}
  \biggr] \ . \nonumber\\
  \label{diquark formula}
\end{eqnarray}
In this expression, the first term in the square brackets is 
the contribution from the quark, and the second term is that from the
antiquark.

The explicit form of the effective potential (\ref{potm}) is given by 
\begin{eqnarray}
 &&\bar{V}_{\rm sol}[\Delta^+ , \Delta^- , B_1 , B_3] \nonumber\\
 &\equiv&{V}[\Delta^+ , \Delta^- , B_1 , B_3]-V[0 , 0 , 0 , 0] \nonumber\\
 &=&-{T}\sum_{n}\int\frac{d^3p}{(2\pi)^3} 
  2\biggl[2\ln\biggl(\frac{F(p,B_1,\Delta)}{[(p_0+\mu)^2-\bar{p}^2]
  [(p_0-\mu)^2-\bar{p}^2]}\biggr) \nonumber\\
 &&\hspace{2.5cm}+\ln\biggl(\frac{F(p,B_3,0)}{[(p_0+\mu)^2-\bar{p}^2]
  [(p_0-\mu)^2-\bar{p}^2]}\biggr)\biggr] \nonumber\\
 &&-{T}\sum_{n}\int\frac{d^3p}{(2\pi)^3} 
  2\biggl[\frac{2}{F(p,B_1,\Delta)}\biggl\{
  [(p_0-\mu)^2-\bar{p}^2-\{B_1(-p)\}^2][(p_0+\mu)^2-\bar{p}^2] \nonumber\\
 &&\hspace{5cm}
  +[(p_0+\mu)^2-\bar{p}^2-\{B_1(p)\}^2][(p_0-\mu)^2-\bar{p}^2] \nonumber\\
 &&\hspace{4cm}-[(p_0)^2-(\bar{p}+\mu)^2]\vert\Delta^-\vert^2
  -[(p_0)^2-(\bar{p}-\mu)^2]\vert\Delta^+\vert^2\biggr\}
  \nonumber\\
 &&\hspace{2cm}+\frac{1}{F(p,B_3,0)}\biggl\{
  [(p_0-\mu)^2-\bar{p}^2-\{B_3(-p)\}^2][(p_0+\mu)^2-\bar{p}^2] \nonumber\\
 &&\hspace{5cm}
  +[(p_0+\mu)^2-\bar{p}^2-\{B_3(p)\}^2][(p_0-\mu)^2-\bar{p}^2]\biggr\}
  \nonumber\\
 &&\hspace{6cm}-6\biggr] \ . 
\label{explicit form of V}
\end{eqnarray}

Substituting the expression for $S_{F11}$
into the SDEs for $B_1$ and $B_3$ in Eqs.~(\ref{SD11a}) and
(\ref{SD11b}), we obtain
\begin{eqnarray}
\label{SDm}
 B_1(p)&=&-{T}\sum_{n=-n_0}^{n_0-1}\int\frac{d\bar{q}\bar{q}^2}{(2\pi)^3}
 2\pi\alpha_s
  \nonumber\\&&\times{K}_0(q_4,p_4,\bar{q},\bar{p})\biggl[
  \frac{5}{6}\frac{F_{+}(q,B_1,\Delta)}{F(q,B_1,\Delta)}
  +\frac{1}{2}\frac{B_3(q)}
  {(iq_4+\mu)^2-\bar{q}^2-\{B_3(q)\}^2}\biggr] \quad , \nonumber\\
\end{eqnarray}
\begin{eqnarray}
\label{SDm3}
 B_3(p)&=&-{T}\sum_{n=-n_0}^{n_0-1}\int\frac{d\bar{q}\bar{q}^2}{(2\pi)^3}
 2\pi\alpha_s
 \nonumber\\&&\times{K}_0(q_4,p_4,\bar{q},\bar{p})\biggl[
  \frac{F_{+}(q,B_1,\Delta)}{F(q,B_1,\Delta)}
  +\frac{1}{3}\frac{B_3(q)}
  {(iq_4+\mu)^2-\bar{q}^2-\{B_3(q)\}^2}\biggr] \quad , \nonumber\\
\end{eqnarray}
where
\begin{eqnarray}
 F_{+}(q,B_1,\Delta)&=&B_1(q)[(iq_4-\mu)^2-\bar{q}^2-\{B_1(-q)\}^2]
  -B_1(-q)\Delta^+(q)\Delta^-(q) \quad , \nonumber\\
\end{eqnarray}
with $q_4=-iq_0=(2n+1)\pi{T}$ and $p_4=-ip_0$. The integration kernel $K_0$ is given by
\begin{eqnarray}
 K_0(q_4,p_4,\bar{q},\bar{p})
  &=&-i\int{d\Omega}D_{\mu\nu}(q-p)
  \mbox{tr}(\gamma^\mu\Lambda_q^\pm\gamma^\nu) \nonumber\\
 &=&\frac{4\pi}{\bar{q}\bar{p}}\log\frac
  {\vert\bar{q}+\bar{p}\vert^2+\omega^2}
  {\vert\bar{q}-\bar{p}\vert^2+\omega^2}
  +\frac{2\pi}{\bar{q}\bar{p}}\log\frac
  {\vert\bar{q}+\bar{p}\vert^2+\omega^2+2M_D^2}
  {\vert\bar{q}-\bar{p}\vert^2+\omega^2+2M_D^2} \quad , \nonumber\\
\end{eqnarray}
where
\begin{eqnarray}
\omega=\vert{q_4-p_4}\vert \ .
\end{eqnarray}
When we set $\Delta^-=\Delta^+=0$, we have 
\begin{eqnarray}
 \frac{F_+(q,B_1,\Delta=0)}{F(q,B_1,\Delta=0)}&=&
  \frac{B_1(q)}{(iq_4+\mu)^2-\bar{q}^2-\{B_1(q)\}^2} \ .
\end{eqnarray}
If we further set $B_1= B_3$, 
the two equations in Eqs.~(\ref{SDm}) and (\ref{SDm3})
become identical.
This implies that $B_1= B_3$ is actually a solution of the SDEs 
for $\Delta^-=\Delta^+=0$.

Next, substituting the expression for $S_{F12}$ into the SDEs 
for $\Delta^-$ and $\Delta^+$ in Eqs.~(\ref{SD12}) and (\ref{SD21}), we obtain
\begin{eqnarray}
\label{SDdn}
 \Delta^-(p)&=&
 {T}\sum_{n=-n_0}^{n_0-1}\int\frac{d\bar{q}\bar{q}^2}{(2\pi)^3}
 2\pi\alpha_s \nonumber\\
 &&\times\biggl[K_1(q_4,p_4,\bar{q},\bar{p})\cdot
  \frac{2}{3}\cdot\frac{G_{+}(q,B_1,\Delta)}{F(q,B_1,\Delta)} 
  +K_2(q_4,p_4,\bar{q},\bar{p})\cdot
  \frac{2}{3}\cdot\frac{G_{-}(q,B_1,\Delta)}{F(q,B_1,\Delta)}\biggr] \ ,
  \nonumber\\
\end{eqnarray}
\begin{eqnarray}
\label{SDdp}
  \Delta^+(p)&=&
  {T}\sum_{n=-n_0}^{n_0-1}\int\frac{d\bar{q}\bar{q}^2}{(2\pi)^3}
  2\pi\alpha_s \nonumber\\
 &&\times\biggl[K_1(q_4,p_4,\bar{q},\bar{p})\cdot
  \frac{2}{3}\cdot\frac{G_{-}(q,B_1,\Delta)}{F(q,B_1,\Delta)} 
  +K_2(q_4,p_4,\bar{q},\bar{p})\cdot
  \frac{2}{3}\cdot\frac{G_{+}(q,B_1,\Delta)}{F(q,B_1,\Delta)}\biggr] \ ,
  \nonumber\\
\end{eqnarray}
where
\begin{eqnarray}
 G_{+}(q,B_1,\Delta)&=&B_1(q)B_1(-q)\Delta^+(q)+[(q_4)^2+(\bar{q}+\mu)^2
  +\vert\Delta^+(q)\vert^2]\Delta^-(q) \ , \nonumber\\ \\
 G_{-}(q,B_1,\Delta)&=&B_1(q)B_1(-q)\Delta^-(q)+[(q_4)^2+(\bar{q}-\mu)^2
  +\vert\Delta^-(q)\vert^2]\Delta^+(q) \ . \nonumber\\
\end{eqnarray}
The integration kernels $K_1$ and $K_2$ are given by
\begin{eqnarray}
 \lefteqn{K_1(q_4,p_4,\bar{q},\bar{p})
  =-i\int{d\Omega}D_{\mu\nu}(q-p)
  \mbox{tr}(\Lambda_p^\pm\gamma^\mu\Lambda_q^\mp\gamma^\nu)} \nonumber\\
  &=&-\frac{\pi(\bar{q}^2-\bar{p}^2)^2}{2\bar{q}^2\bar{p}^2\omega^2}
  \ln\frac{(\bar{q}+\bar{p})^2}{(\bar{q}-\bar{p})^2}
  +\frac{\pi}{2\bar{q}^2\bar{p}^2}\biggl\{
  \frac{(\bar{q}^2-\bar{p}^2)^2}{\omega^2}+4\bar{q}\bar{p}-\omega^2
  \biggr\}
  \ln\frac{(\bar{q}+\bar{p})^2+\omega^2}{(\bar{q}-\bar{p})^2+\omega^2}
  \nonumber\\
 &&+\frac{\pi}{2\bar{q}^2\bar{p}^2}\biggl[
  \frac{(\bar{q}^2-\bar{p}^2)^2}{2M_D^2+\omega^2}
  \ln\frac{(\bar{q}+\bar{p})^2}{(\bar{q}-\bar{p})^2} \nonumber\\
 &&\hspace{0.5cm}+\frac{\{(\bar{q}+\bar{p})^2+2M_D^2+\omega^2\}
  \{(2M_D^2+\omega^2)^2+(\bar{q}-\bar{p})^2\omega^2\}}
  {2M_D^2(2M_D^2+\omega^2)}\ln\frac{(\bar{q}+\bar{p})^2+2M_D^2+\omega^2}
  {(\bar{q}-\bar{p})^2+2M_D^2+\omega^2} \nonumber\\ 
 &&\hspace{0.5cm}
  -\frac{\{(\bar{q}+\bar{p})^2+\omega^2\}\{(\bar{q}-\bar{p})^2+\omega^2\}}
  {2M_D^2}\ln\frac{(\bar{q}+\bar{p})^2+\omega^2}
  {(\bar{q}-\bar{p})^2+\omega^2}
  \biggr] \ , 
\end{eqnarray}
\begin{eqnarray}
 \lefteqn{K_2(q_4,p_4,\bar{q},\bar{p})
  =-i\int{d\Omega}D_{F\mu\nu}(q-p)
  \mbox{tr}(\Lambda_p^\pm\gamma^\mu\Lambda_q^\pm\gamma^\nu)} \nonumber\\
 &=&\frac{\pi(\bar{q}^2-\bar{p}^2)^2}{2\bar{q}^2\bar{p}^2\omega^2}
  \ln\frac{(\bar{q}+\bar{p})^2}{(\bar{q}-\bar{p})^2}
  -\frac{\pi}{2\bar{q}^2\bar{p}^2}\biggl\{
  \frac{(\bar{q}^2-\bar{p}^2)^2}{\omega^2}-4\bar{q}\bar{p}-\omega^2
  \biggr\}
  \ln\frac{(\bar{q}+\bar{p})^2+\omega^2}{(\bar{q}-\bar{p})^2+\omega^2}
  \nonumber\\
 &&-\frac{\pi}{2\bar{q}^2\bar{p}^2}\biggl[
  \frac{(\bar{q}^2-\bar{p}^2)^2}{2M_D^2+\omega^2}
  \ln\frac{(\bar{q}+\bar{p})^2}{(\bar{q}-\bar{p})^2} \nonumber\\
 &&\hspace{0.5cm}+\frac{\{(\bar{q}-\bar{p})^2+2M_D^2+\omega^2\}
  \{(2M_D^2+\omega^2)^2+(\bar{q}+\bar{p})^2\omega^2\}}
  {2M_D^2(2M_D^2+\omega^2)}\ln\frac{(\bar{q}+\bar{p})^2+2M_D^2+\omega^2}
  {(\bar{q}-\bar{p})^2+2M_D^2+\omega^2} \nonumber\\ 
 &&\hspace{0.5cm}
  -\frac{\{(\bar{q}+\bar{p})^2+\omega^2\}\{(\bar{q}-\bar{p})^2+\omega^2\}}
  {2M_D^2}\ln\frac{(\bar{q}+\bar{p})^2+\omega^2}
  {(\bar{q}-\bar{p})^2+\omega^2}
  \biggr] \ . 
\end{eqnarray}
We note that for $B_1=0$, we have
\begin{eqnarray}
 \frac{G_{\pm}(q,B_1=0,\Delta)}{F(q,B_1=0,\Delta)}&=&
  \frac{-\Delta^\mp(q)}{(q_4)^2+(\bar{q}\mp\mu)^2+\vert\Delta^\mp(q)\vert^2}
  \ . 
\end{eqnarray}
Thus, the SDEs (\ref{SDdn}) and (\ref{SDdp}) are identical to 
the well-known
forms given in,~e.g.,~Ref.~\citen{Hong}.

\end{document}